\documentclass[reprint,english,superscriptaddress,pra]{revtex4-2}

\usepackage[latin1]{inputenc}
\usepackage{graphicx}
\usepackage{amsmath}
\usepackage{amsthm}
\usepackage{bm}
\usepackage{layout}
\usepackage{float}
\usepackage{amsfonts}
\usepackage{amssymb}%
\usepackage[margin=0.75in]{geometry}
\usepackage{color}
\usepackage{soul}
\usepackage{mathtools}
\usepackage[colorlinks=true,citecolor=blue]{hyperref}
\theoremstyle{definition}

\usepackage[capitalise,compress]{cleveref}

\newcommand{\ket}[1]{\left | #1 \right\rangle}

\newcommand{\bra}[1]{\left \langle #1 \right |}

\newcommand{\abs}[1]{\left | #1 \right|}
\renewcommand{\epsilon}{\varepsilon}
\renewcommand{\O}[1]{ O\left(#1\right)}

\newcommand{\norm}[1]{\left\|#1\right\|}

\newcounter{para}
\newcommand{\dist}{\textrm{dist}}

\makeatletter
\newcommand*\bigcdot{\mathpalette\bigcdot@{.5}}
\newcommand*\bigcdot@[2]{\mathbin{\vcenter{\hbox{\scalebox{#2}{$\m@th#1\bullet$}}}}}

\makeatother

\usepackage{array}
\newcolumntype{L}{>{$}l<{$}} 
\newcolumntype{C}{>{$}c<{$}} 
\newcolumntype{R}{>{$}r<{$}} 

\usepackage{xcolor}

\usepackage{mathtools}

\usepackage{xr}
\makeatletter
\newcommand*{\addFileDependency}[1]{
  \typeout{(#1)}
  \@addtofilelist{#1}
  \IfFileExists{#1}{}{\typeout{No file #1.}}
}
\makeatother

\usepackage{tikz}

\usepackage{mdframed}
\newmdenv[topline=false,rightline=false,bottomline=false,linewidth=2pt,linecolor=white!60!black,]{leftborder}

\usepackage{qcircuit}

\usepackage{booktabs}
\newcolumntype{C}{>{$}c<{$}}
\AtBeginDocument{
\heavyrulewidth=.08em
\lightrulewidth=.05em
\cmidrulewidth=.03em
\belowrulesep=.65ex
\belowbottomsep=0pt
\aboverulesep=.4ex
\abovetopsep=0pt
\cmidrulesep=\doublerulesep
\cmidrulekern=.5em
\defaultaddspace=0.5em
}
\newcommand{\GHZ}{\ket{\text{GHZ}}}
\newcommand{\GHZab}{\ket{\text{GHZ}(a,b)}}

\newcommand{\polylog}{\text{polylog}}

\usepackage[english]{babel}
\makeatletter
\let\ORIbbl@fixname\bbl@fixname
\def\bbl@fixname#1{%
  \@ifundefined{languagealias@\expandafter\string#1}
    {\ORIbbl@fixname#1}
    {\edef\languagename{\@nameuse{languagealias@#1}}}%
}
\newcommand{\definelanguagealias}[2]{%
  \@namedef{languagealias@#1}{#2}%
}
\makeatother

\definelanguagealias{en}{english}
\definelanguagealias{EN}{english}

\usepackage{amstext} 
\usepackage{array}   
\newcolumntype{L}{>{$}l<{$}} 

\usepackage{subfigure,physics}
\newcommand{\eps}{\ensuremath{\varepsilon}}

\newcommand{\al}{\ensuremath{\alpha}}

\DeclarePairedDelimiter\parentheses{\lparen}{\rparen}
\renewcommand{\O}[1]{\mathcal{O}\parentheses*{#1}}
\newcommand{\W}[1]{\Omega\parentheses*{#1}}

\newcommand{\etal}{\emph{et al.}~}

\crefname{section}{Sec.}{Secs.}
\crefname{section}{Section}{Sections}
\crefrangelabelformat{equation}{\textup{(#3#1#4)}--\textup{(#5#2#6)}}

\begin{document}

\title{Experimental roadmap for optimal state transfer and entanglement generation in power-law systems}
\author{Andrew~Y.~Guo}
\email[Corresponding author: ]{guoa@umd.edu}
\affiliation{Joint Center for Quantum Information and Computer Science, NIST/University of Maryland, College Park, MD 20742, USA}
\affiliation{Joint Quantum Institute, NIST/University of Maryland, College Park, MD 20742, USA}

\author{Jeremy~T.~Young}
\thanks{The first two authors contributed equally}
\affiliation{Institute of Physics, University of Amsterdam, 1098 XH Amsterdam, the Netherlands}
\affiliation{JILA, University of Colorado and National Institute of Standards and Technology, and Department of Physics, University of Colorado, Boulder, Colorado 80309, USA}
\affiliation{Center for Theory of Quantum Matter, University of Colorado, Boulder, Colorado 80309, USA}

\author{Ron Belyansky}
\affiliation{Joint Center for Quantum Information and Computer Science, NIST/University of Maryland, College Park, MD 20742, USA}
\affiliation{Joint Quantum Institute, NIST/University of Maryland, College Park, MD 20742, USA}

\author{Przemyslaw~Bienias}
\affiliation{Joint Center for Quantum Information and Computer Science, NIST/University of Maryland, College Park, MD 20742, USA}
\affiliation{Joint Quantum Institute, NIST/University of Maryland, College Park, MD 20742, USA}

\author{Alexey~V.~Gorshkov}
\affiliation{Joint Center for Quantum Information and Computer Science,
NIST/University of Maryland, College Park, MD 20742, USA}
\affiliation{Joint Quantum Institute, NIST/University of Maryland, College Park, MD 20742, USA}

\date{\today}

\begin{abstract}
	Experimental systems with power-law interactions have recently garnered interest as promising platforms for quantum information processing. Such systems are capable of spreading entanglement superballistically and achieving an asymptotic speed-up over locally interacting systems.
    Recently, protocols developed by Eldredge \etal[Phys.\ Rev.\ Lett.\ \textbf{119}, 170503 (2017)] and Tran \etal[Phys.\ Rev.\ X \textbf{11}, 031016 (2021)] for the task of transferring a quantum state between distant particles quickly were shown to be optimal and saturate theoretical bounds.
    However, the implementation of these protocols in physical systems with long-range interactions remains to be fully realized.
    In this work, we provide an experimental roadmap towards realizing fast state-transfer protocols in three classes of atomic and molecular systems with dipolar interactions: polar molecules composed of alkali-metal dimers, neutral atoms in excited Rydberg states, and atoms with strong magnetic moments (e.g. dysprosium).  As a guide to near-term experimental implementation, we numerically evaluate the tradeoffs between the two protocols for small system sizes and develop methods to address potential crosstalk errors that may arise during the execution of the protocols.
\end{abstract}

\maketitle

Ultracold quantum systems have recently attracted attention for their potential for quantum information processing, and innovations in the individual control of ultracold atoms \cite{Endres2016,Barredo2018,Browaeys2020,Kaufman2021,Bloch2023,Fraxanet2022,Gross2021,Anich2023,Su2023,Sohmen2023, Chomaz2022} and molecules \cite{Liu2018,Anderegg2019,Zhang2022, Vilas2023,Covey2018,Christakis2023,Rosenberg2022,Tobias2022} have led to the realization of high-fidelity quantum gates \cite{Levine2018,Levine2019,Graham2019,Madjarov2020,Schine2022,Ma2023,Bluvstein2023,Holland2023,Bao2023}.
Many atomic and molecular systems also have access to long-range interactions, which can generate entanglement
quickly between qubits at large distances. 
Compared to finite-range couplings, long-range interactions allow for increased connectivity between qubits, which can lead to speed-ups for performing general computational tasks \cite{Chen2023}.

In particular, it was recently shown that systems with power-law interactions---i.e. those that decay as $1/r^\al$ in the distance between qubits---can achieve an asymptotic speed-up for\emph{quantum state transfer}, a task wherein a quantum state must be transferred quickly between two distant qubits \cite{Eldredge2017,Tran2021a,Hong2021}.
Performance on the state transfer task provides a benchmark for a system's ability to execute nonlocal quantum gates quickly, and also serves as a computational primitive for more complicated circuit operations, such as quantum routing \cite{Bapat2023}.
Additionally, many state-transfer protocols often generate many-body entanglement as an intermediate step.
The generation of such many-body entangled states as the  Greenberger-Horne-Zeilinger (GHZ) and W states can be an important computational resource in itself for applications such as sensing at the Heisenberg limit \cite{Bollinger1996,Eldredge2018} and performing multi-qubit quantum gates \cite{Guo2022}.

In addition to their usefulness in generating entanglement quickly, state-transfer protocols have also attracted significant theoretical interest due to their ability to saturate fundamental Lieb-Robinson bounds on the rate at which information can propagate in quantum systems \cite{Hastings2006,Guo2020,Tran2019,kuwaharaStrictlyLinearLight2020,Tran2021b,Chen2023}.
In particular, the state-transfer protocols in Refs.~\cite{Eldredge2017} and \cite{Tran2021a} have been shown to saturate the best Lieb-Robinson bounds for power-law interactions \cite{Tran2021b}.
In addition to being asymptotically optimal, these fast entangling protocols imply that power-law interactions are qualitatively stronger than previously understood; indeed, while state transfer in faster-than-linear time has been shown before using interactions with relatively small power-law exponents (e.g.~$\al\approx 1$ in one-dimensional [1D] systems \cite{Richerme2014}), realizing the same state transfer times using interactions with more rapid power-law decays (e.g. van der Waals interactions with $\al=6$ in 3D) would be the first demonstration of fast information transfer using interactions previously thought to be effectively short-range.
\begin{figure*}[t]
	\subfigure[]{
		\includegraphics[width = 0.25\textwidth]{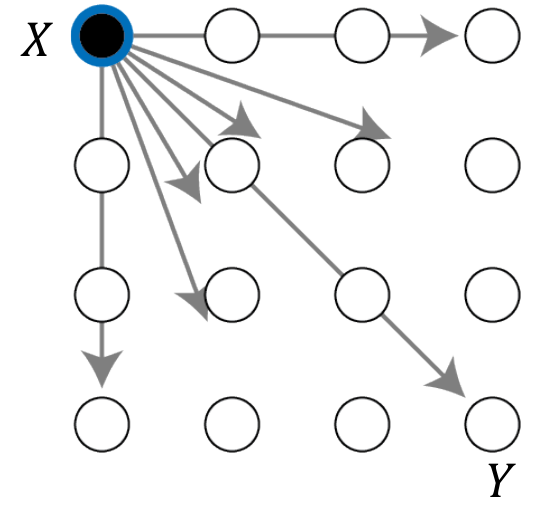}
	}
 \qquad \qquad \qquad \qquad \qquad
	\subfigure[]{
		\includegraphics[width = 0.25\textwidth]{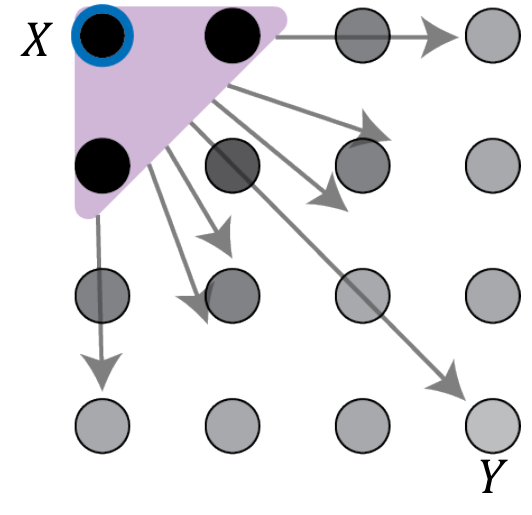}
	}
	\caption{Schematic for the Eldredge state-transfer protocol. (a) Interactions from $H(t)$ fan out the state from the initial qubit $X$ (black) to the rest of the lattice. (b) Once the $\textsc{CNot}$ gate has been completed on a ``target'' qubit, it becomes a ``control'' qubit and part of the GHZ-like state (purple shaded region). The remaining, ``target'' qubits are shaded gray according to the phase that has accumulated due to the interacting Hamiltonian and will be added to the GHZ state in future steps of the protocol. Afterwards, the GHZ-state creation is reversed and the state is transferred to the final qubit, $Y$.}
	\label{fig:Zach}
\end{figure*}

In this work, we present a roadmap for the experimental implementation of fast quantum state transfer protocols using long-range interactions.
We first review the power-law interacting protocols  developed by Eldredge \etal\cite{Eldredge2017} and Tran \etal\cite{Tran2021a} in \cref{sec:state-transfer-protocols}, and analyze both their asymptotic runtimes as well as their numerical performance for finite system sizes.
Then, in \cref{sec:expt_realization}, we provide schemes for realizing these protocols in three model power-law interacting systems: polar molecules, Rydberg atoms, and atoms with large magnetic moments (dysprosium in particular).
Our approach utilizes only global power-law interactions and single-qubit control.
In \cref{sec:crosstalk}, we analyze the contribution of certain forms of crosstalk errors that may arise in the course of the state transfer and propose a few methods to mitigate those errors.

\section{Fast state-transfer protocols for power-law interactions}
\label{sec:state-transfer-protocols}

In this section, we will review the two protocols for performing fast quantum state transfer using power-law interactions developed by Eldredge \etal \cite{Eldredge2017} and Tran \etal \cite{Tran2021a}.
We will summarize the protocols briefly, present the asymptotic scaling of their state transfer times, and compare their relative advantages in finite-size systems.

The state transfer task is given as follows.
Given a system of $N$ qubits, we wish to transfer an unknown input state $\ket{\psi} = a \ket{0} + b\ket{1}$ initially located on qubit $X$ to qubit $Y$.
We assume that the intermediate qubits (as well as qubit $Y$) are initialized in state $\ket{0}$.
Then the state transfer task is to realize the following unitary $U$:
\begin{align}
	U\bigl(\ket{\psi}_X\ket{0}^{N-2}\ket{0}_Y\bigr) \mapsto \ket{0}_X\ket{0}^{N-2}\ket{\psi}_Y.
\end{align}
For the rest of this paper, we will assume that the system is arranged in a $d$-dimensional grid of qubits of linear size $L \equiv \sqrt[d]{N}$, although other lattice configurations are possible.
The state transfer will take place over a distance $r = \text{dist}(X,Y)$, which can be as large as $r_\text{max} = L\sqrt{d}$.

The Eldredge protocol achieves state transfer by first encoding the qubit state into a many-body entangled state that includes the initial and target qubits (cf.~\cref{fig:Zach}).
It does so by performing a sequence of cascaded controlled-$\textsc{Not}$ ($\textsc{CNot}$) gates, which realize the operation
\begin{equation}
	\textsc{CNot } (a\ket{0} + b\ket{1})\ket{0} = a\ket{00} + b\ket{11}.
\end{equation}
The $\textsc{CNot}$ gate from qubit $i$ to qubit $j$ can be implemented by a Hamiltonian $H_{ij} = h_{ij} \ketbra{1}_i\otimes X_j$ acting for time $t = \pi/(2h_{ij})$, up to a local unitary term.
In Ref.~\cite{Eldredge2017}, a Hamiltonian $H(t) = \sum_{ij}H_{ij}(t)$ is applied that progressively turns on/off the interactions between pairs of qubits at different times and enables the encoding of the state $\ket{\psi}$ into a GHZ-like state $\ket{\text{GHZ}(a,b)} = a \ket{0}^{\otimes N} + b \ket{1}^{\otimes N}$.
The protocol starts by acting on all qubits in the lattice using a single control qubit storing the initial state. At each time step, once the $\textsc{CNot}$ is fully complete on a target qubit, then it is added to the list of controls.
At each point in time, multiple controls may act on a single target.
This process continues until all of the qubits have been added to the GHZ state.
The time required to create the full GHZ state is the sum of these time steps.
At the end of the process, the protocol is reversed, interchanging the roles of qubits $X$ and $Y$, to implement the full state transfer.

By using a long-range interacting Hamiltonian, the protocol developed by Eldredge \etal is able to achieve a sublinear state transfer time.
Assuming that $h_{ij} = 1/r_{ij}^\al$, where $r_{ij}$ is the distance between qubits $i$ and $j$, the time required to complete the $\textsc{CNot}$ on qubit $j$ being acted on by controls indexed by $i$ is equal to
\begin{align}
	t = \frac{\pi}{2\sum_i h_{ij}} = \frac{\pi}{2\sum_ir^{-\al}_{ij}}.
 \end{align}
The scaling of the state transfer time $t_\text{ST}$ using the protocol derived in Ref.~\cite{Eldredge2017} is given by
\begin{equation}
\label{eq:statetransfertimes_Zach}
	t_\text{ST} =
  \begin{cases}
    \O{1} & \alpha < d
    \\ \O{\log r} & \alpha = d
    \\ \O{r^{\min(\al-d,1)}} & \alpha > d.
  \end{cases}
\end{equation}
The intuition for the various scalings are as follows:
for $\al < d$, the extreme non-locality of the interactions causes the state transfer velocity to diverge in the thermodynamic limit, which implies that the state can be transferred over arbitrary distances in constant time, so the scaling of $t_\text{ST}$ is independent of $r$.

For $\al \in [d,d+1]$, take a block of qubits of linear size $r$ and suppose we wish to double $r$.
Then the qubit at the corner of the next block is acted on by controlled rotations of total strength proportional to $\{\text{\# of controls acting on the qubit\}}/r^\al \propto r^{d-\al}$.
Thus, for $\al \in (d,d+1)$, the state-transfer time is proportional to $r^{\al-d}$, which is sublinear. 

For $\al = d$, the time it takes to double the linear block size is constant, so it takes overall $\log r$ time to implement the state transfer protocol.
Lastly, for $\al = d+1$, the state-transfer time is linear in $r$, thus recovering the nearest-neighbor-interaction limit. 

For $\al > d+1$, the doubling approach thus leads to worse scaling than nearest-neighbor interactions. Instead, the incremental approach which immediately converts targets into controls must be used, and there is at most a constant speedup from the power-law interactions. This incremental approach realizes the same scaling as the doubling approach for $d \leq \alpha \leq d+1$ and is the optimal form of the Eldredge protocol for isotropic power-law interactions.
In conclusion, the Eldredge protocol yields an asymptotic speed-up over nearest-neighbor interactions for all $\al < d+1$.

\begin{figure}[t]
	\includegraphics[width = 0.475\textwidth]{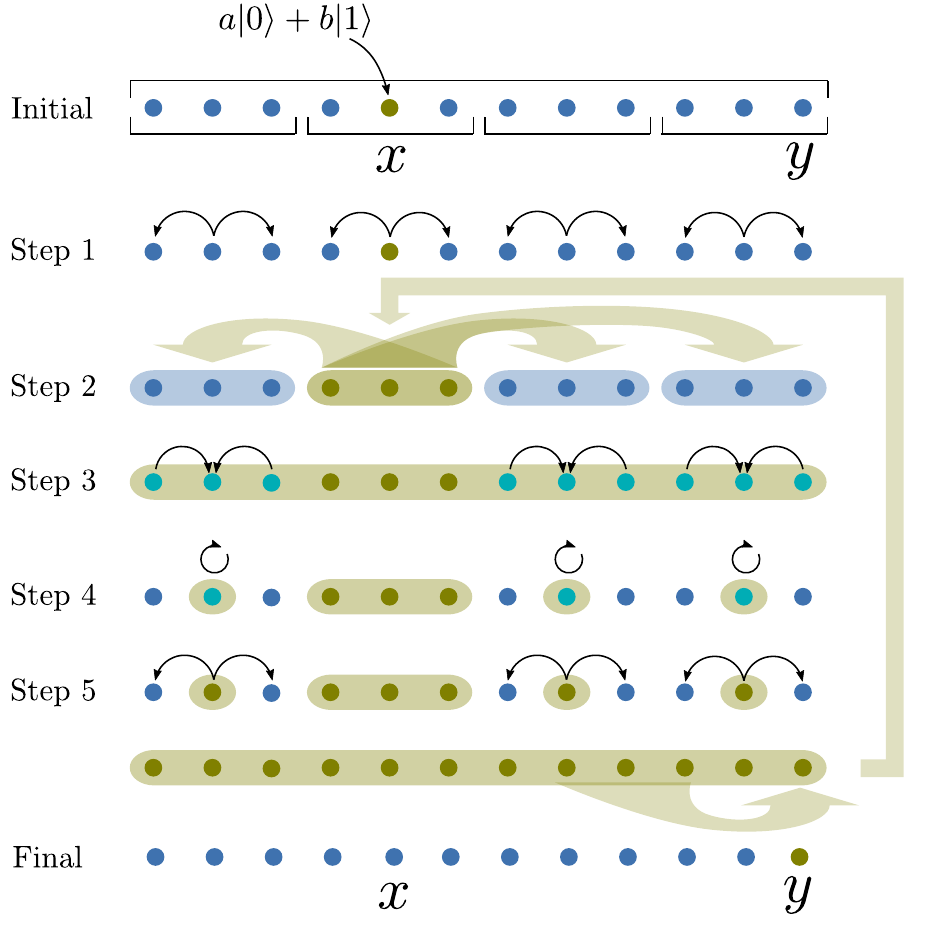}
	\caption{Outline of the Tran protocol. We start with qubit $X$ in an unknown state $\ket{\psi} = a \ket{0} + b\ket{1}$, while the rest of the qubits start in state $(\ket{0} + \ket{1})/\sqrt{2}$.
 In step 1, either nearest-neighbor interactions or the Eldredge protocol are used to create a GHZ state on neighboring qubits. Then, in step 2, controlled-phase interactions are used to merge together neighboring GHZ states. The last three steps (steps 3-5) serve to transform the state into the desired GHZ-like state by moving the entanglement into a single qubit, performing a single-qubit rotation, and then reentangling the qubits. This final GHZ-like state is fed back into the protocol recursively, after which the protocol is implemented in ``reverse'' in order to implement the state transfer from qubit $X$ onto qubit $Y$.}
	\label{fig:Tran}
\end{figure}

The protocol discovered by Tran \etal is able to achieve an even faster speed-up for the state transfer task (see \cref{fig:Tran}).
As with the Eldredge protocol, it uses a GHZ state as an intermediate and performs the state transfer recursively by encoding the qubit state $\ket{\psi}$ into progressively larger GHZ-like states until the state includes the whole sublattice of sites between site $X$ and site $Y$ \cite{Tran2021a}.
Then the protocol reverses the GHZ-state encoding procedure to map the state of the original qubit on site $X$ onto the qubit at site $Y$.
In the process, small GHZ-like states are merged together into larger GHZ-like states.
The merge subroutine is performed using diagonal interactions of the form $H_{ij} = h_{ij} \ketbra{1}_i \otimes \ketbra{1}_j$, which serve to perform a controlled-PHASE gate between qubits $i$ and $j$.

The scaling of the state transfer time $t_\text{ST}$ of the Tran protocol is given as follows:
\begin{equation}
\label{eq:statetransfertimes_Minh}
	t_\text{ST} =
  \begin{cases}
	\O{\log^{\kappa_\alpha} r }& \alpha \in (d,2d)
    \\ \O{e^{\gamma\sqrt{d \log r}}}& \al = 2d
    \\ \O{r^{\min(\al - 2d,1)}} & \alpha > 2d.
  \end{cases}
\end{equation}
Due to the protocol's recursive nature, it can transfer a state in time that scales faster than any polynomial in $r$ for all $\al < 2d$.
For $\al = 2d$, the state-transfer time is superlogarithmic---but still subpolynomial---and for $\al \in (2d, 2d+1)$, the time is sublinear and proportional to $r^{\al-2d}$.
The intuition for the latter scaling is as follows: consider two blocks of GHZ-like states of linear size $r$ situated a distance $r$ apart.
The time required to merge the two blocks scales inversely with the combined interaction strength, which is proportional to $\{\text{\# of interacting pairs of qubits\}}/r^\al \propto r^{2d-\al}$.
For more details on the implementation of the protocol, see \cref{app:ost-protocol}.

Thus, for all $\al < 2d+1$, the Tran protocol allows for even more significant speed-ups over nearest-neighbor systems than the Eldredge protocol.
This regime of $\alpha$ includes  dipolar interactions in 2D and 3D and van der Waals interactions in 3D. For dipolar interactions in 3D, both the Eldredge and Tran protocols realize exponential speedups. However, since the Eldredge protocol still retains a polynomial speedup over the Tran protocol for $\al = d$, in the rest of the paper, we shall focus on dipolar interactions only in 2D.

\subsection{Comparison of state-transfer protocols}
\label{subsec:comparison}


\begin{figure*}
	\subfigure{
	\label{fig:crossover_point}
	\includegraphics[width = 0.45\textwidth]{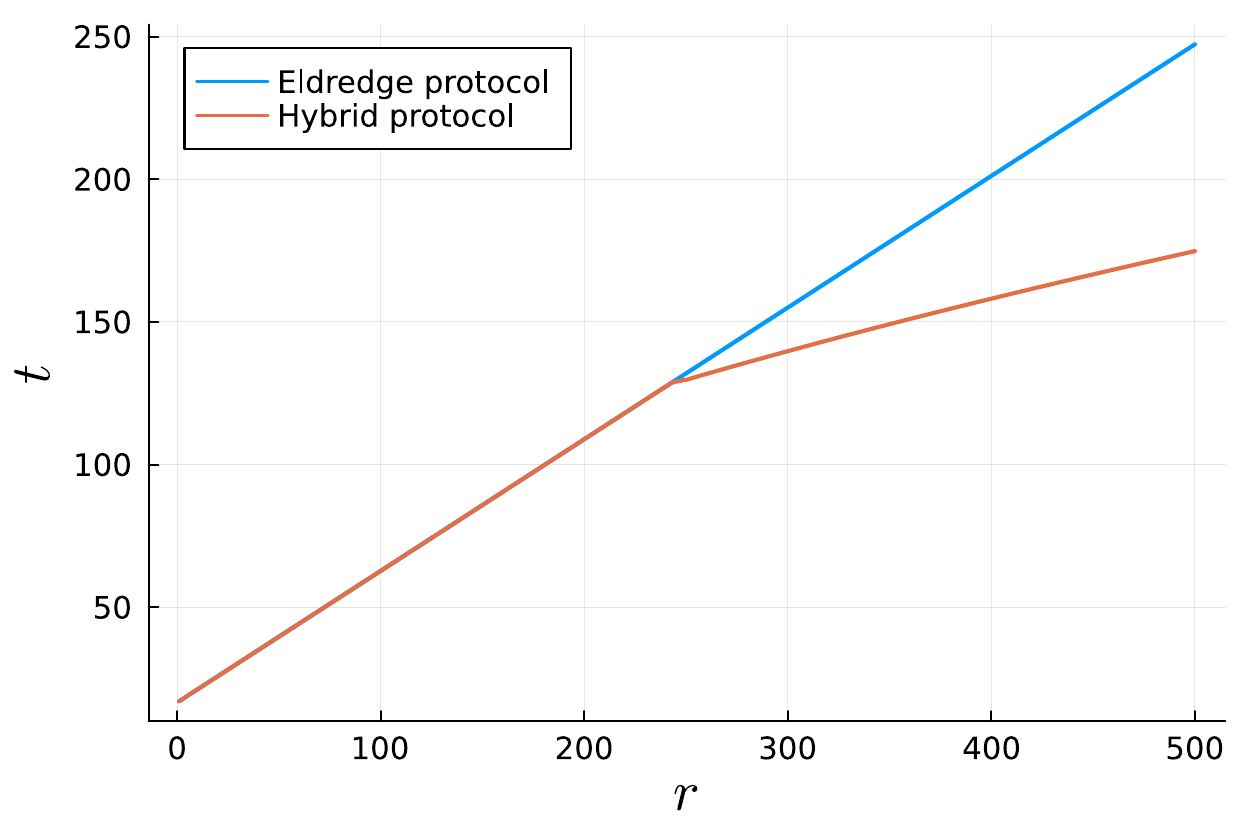}
	}
 \qquad
	\subfigure{
	\label{fig:m_scaling}
	\includegraphics[width = 0.45\textwidth]{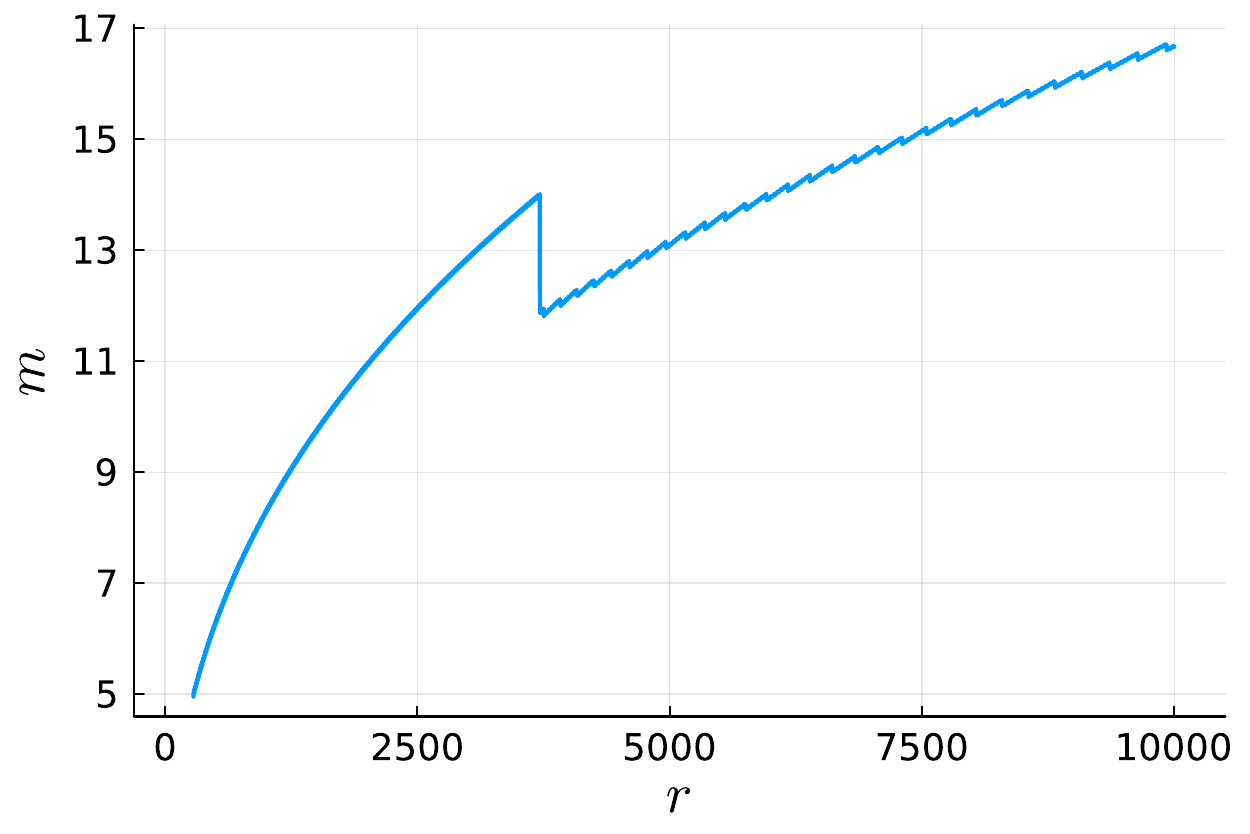}
	}

	\caption{(a) Comparison of the scaling of the state-transfer times for the two protocols described in \cref{subsec:comparison} as a function of the characteristic length $r$, using dipolar interactions in 2D ($\al = 3$ and $d = 2$).
    The  hybrid protocol consists of a single layer of the recursive Tran protocol on top of the base Eldredge protocol, and the crossover point where it improves over the pure Eldredge protocol occurs at a characteristic length of $r_* = 244$.
    (b) The scaling of the optimal values of $m = r/r_1$ for the hybrid protocol as a function of $r$, as determined numerically using a dynamic programming algorithm.
    The curve begins at the characteristic length $r_* = 244$, when the first layer of the Tran protocol is applied. The stepwise transition for $r \approx 3700$ indicates the transition from a single recursive step to two. The jaggedness in the curve arises due to the fact that the $m$ values scale inversely with $r_1$, which can only take on discrete values by assumption. }
	\label{fig:Zach_vs_Minh}
\end{figure*}

While the state-transfer protocol developed by Tran \etal in Ref.~\cite{Tran2021a} is asymptotically faster than the one presented by Eldredge \etal in Ref.~\cite{Eldredge2017} for all $\al > d$, it is less efficient to implement for small system sizes.
This is due to the recursive step needing to  call a subroutine that creates smaller GHZ states a total of three times.
Instead of using the Tran protocol for small $r$, it would be faster to use the Eldredge protocol for $r$ up to a fixed size, and then utilize the recursive step of the former protocol to achieve larger GHZ states.
In the rest of this subsection, we numerically determine the optimal crossover point when a hybrid protocol consisting of performing the Eldredge protocol followed by a single step of the Tran protocol surpasses the pure Eldredge protocol for dipole-dipole interactions on a 2D lattice.

To implement the Eldredge protocol described in \cref{sec:state-transfer-protocols}, specifically the incremental approach which immediately converts target qubits into control qubits, we numerically calculate the state transfer times for system sizes up to $r \le 110$, and then use a linear fit to extrapolate to larger values of $r$ (since the asymptotic runtime of the protocol was shown to obey the scaling $t \propto r$ for $\al \ge d+1$). We use these state transfer times as the benchmark to compare the hybrid protocol.

To implement the hybrid protocol, we apply a recursive approach known as \emph{bottom-up dynamic programming}: first, we choose a fixed value of $r=r_0$ below which we will use the Eldredge protocol to create a GHZ state of that size.
Then, for all larger $r$, we recursively calculate the time $t(r;r_1)$ required to create a GHZ state of size $r$ by assembling it from smaller GHZ states of size $r_1 < r$.
By minimizing $t(r;r_1)$ over all values of $r_1$,  we obtain the optimal creation time $t(r)$ for a GHZ state of size $r$ using a single step of the Tran protocol to get from $r_1$ to $r$.
We note that this protocol could require multiple recursive steps, since creating a GHZ-state of size $r_1$ could itself  potentially require recursive steps, thus leading to the need to apply multiple layers of the Tran protocol.

In \cref{fig:crossover_point}, we show the transition point above which a single step of the hybrid protocol can create a GHZ state faster than the pure Eldredge protocol.
For $\al = 3$ on a 2D lattice, this critical point occurs at $r_* = 244$, which corresponds to a lattice size of $N = r_*^2 = 59,538$.
We note that this is an estimate of the true crossover point, since we did not exactly calculate the exact state transfer time for the pure Eldredge protocol beyond $r = 110$ due to the numerical cost of the algorithm (with runtime scaling as $\O{r^6}$).

In \cref{fig:m_scaling}, we show the scaling of the optimal choice of the ratio of the sizes of the GHZ states in the recursive step of the protocol, which we define to be equal to $m \equiv r/r_1$ for a given choice of $r$.
For one step of the hybrid protocol, $m$ ranges from 5 to 13.
Asymptotically, the analysis of the runtime of the protocol predicts that $m$ will scale as $m \propto r^{2d/\al - 1} = r^{1/3}$ \cite{Tran2021a}.

In order to recover the optimal asymptotic scaling of the state-transfer time for the hybrid protocol, it will in general be necessary to use the recursive step of the hybrid protocol multiple times.
For a fixed $r_0$, the number of times the recursive step is called will scale as $n = \O{\log\log(r)}$, as one can show by applying the relation $r_1 = r/m = r^{2/3}$ recursively $n$ times to get the equation $(r^{2/3})^n = r_0$ and solving for $n$.

In summary, using a hybrid of both the Tran and Eldredge protocols to create an intermediate GHZ state allows for faster state-transfer times than using either of those protocols alone.
Our estimate of the crossover point suggests that the performance of the hybrid protocol surpasses that of the Eldredge protocol for system sizes exceeding roughly 60,000 qubits.
As such, the speed-up for the hybrid protocol could be potentially be realized in larger-scale quantum devices.

\section{Proposal for experimental realization of state-transfer protocols}
\label{sec:expt_realization}

Before explaining the details of how to implement the state-transfer protocol for particular systems, we will first discuss the features that apply to all of the following systems.

In the state-transfer protocol, one generically has three types of qubits: control qubits, target qubits, and uninvolved qubits. The uninvolved qubits can generally be prepared by storing them in states (e.g., two hyperfine ground states for neutral atoms) that do not interact with the other two types of qubit or each other, which we discuss in further detail below. For the control and target qubits, we must engineer a Hamiltonian with long-range Ising interactions between control and target qubits. However, the control qubits must not interact with themselves, and similarly the target qubits must not interact with themselves, neither of which occurs generically, so these interactions must be engineered.

To do so, we adopt the spin-echo-like approach from Ref.~\cite{Eldredge2017}. This approach relies on the ability to realize a Hamiltonian of the form
\begin{equation}
    H_\text{int} = \frac{1}{2} \sum_{i \neq j} J_{ij} Z_i Z_j,
\end{equation}
as well as its negative $H'_{\text{int}}\equiv-a H_\text{int}$, where $Z_i$ denotes the Pauli-Z matrix for the $i$th qubit and $a > 0$. By evolving under $H_\text{int}$ for a time $t$, applying a Pauli-X gate to either the control qubits or the target qubits, and evolving under $-a H_\text{int}$ for a time $t/a$, followed by another Pauli-$X$ gate on the same set of qubits, the evolution from the control-control and target-target interactions is undone while the control-target interactions remain, as desired. In light of this, most of the following discussion will be centered around how to realize both signs of the Ising interaction. Often, the engineered interactions include an inhomogeneous longitudinal field term. The corresponding evolution due to this term can be easily corrected via single-qubit gates.

The necessary local control can be achieved in tweezer arrays of Rydberg atoms
\cite{Endres2016,Barredo2018,Browaeys2020,Kaufman2021},
polar molecules \cite{Liu2018,Anderegg2019,Zhang2022,Vilas2023}, and very recently dysprosium \cite{Bloch2023} or by using quantum gas microscopes for polar molecules \cite{Covey2018,Christakis2023,Rosenberg2022,Tobias2022} and dysprosium \cite{Fraxanet2022,Gross2021,Anich2023}. As far as magnetic atoms go, we will focus on dysprosium in this paper, but similar advances have been made for quantum gas microscopes of erbium \cite{Su2023} as well as for mixtures of dysprosium and erbium \cite{Sohmen2023}. Additionally, when utilizing $1/r^3$ dipole-dipole interactions to realize the desired Ising interactions, they will be proportional to $1 - 3 \cos^2 \theta_{ij}$, where $\theta_{ij}$ is the polar angle of $\mathbf{r}_i - \mathbf{r}_j$ for particles $i$ and $j$ relative to the quantization axis. Since the protocol relies on all interactions having the same sign, for dipole-dipole interactions we shall restrict our focus to 2D with $\theta = \pi/2$. Moreover, the Eldredge protocol can already be utilized with an exponential speedup in 3D. In the case of Rydberg atoms, isotropic vdW interactions can emerge, so we also discuss how to realize the desired positive and negative interactions for this case in 3D.

Finally, we note that it is important to be able to turn off the interactions for individual qubits as necessary.
There are two main ways that this can be done.
First, the qubit states can be transferred to non-interacting states. For example, when considering Rydberg systems, the Rydberg states can be coherently transferred to non-interacting ground states via an optical drive. 
In some cases, there may not be any fully non-interacting states which can be utilized during the protocol.
In these cases, one can utilize similar spin-echo techniques as discussed above in order to remove the unwanted interactions. This can be achieved as follows. First, during the positive interaction evolution, each target qubit requires an interaction time $t_i$. However, without non-interacting shelving states, each control qubit will interact with the target qubits for a time $t_\text{max} \equiv \max_i t_i$. To remove the excess interactions felt by an individual target qubit, a $\pi$ pulse is applied to the corresponding target qubit after time $t_i + (t_\text{max}-t_i)/2$, and the remaining evolution time of $(t_\text{max}-t_i)/2$ reverses the interactions, leading to an effective interaction time of $t_i$. The same approach is used for the negative interaction evolution, thus removing all unwanted interactions.

\subsection{Polar molecules}
\label{subsec:polar-molecules}

In the following subsections, we provide further details regarding the experimental implementation of the state-transfer procedures described in \cref{sec:state-transfer-protocols}.
First, we will describe how to perform state transfer using dipole-dipole interactions in a system of polar alkali-metal dimers.

We will assume the molecules are arrayed in a 2D lattice with a single molecule per lattice site and negligible tunneling of molecules.
The single molecule Hamiltonian is
\begin{equation}
    H_0 = B_0 \mathbf{N}^2 - d_0 E,
\end{equation}
where $B_0$ is the rotational constant of the electronic and vibrational ground state of the molecule, $E$ is the electric field strength, and $d_p = \mathbf{\hat{e}_p} \cdot \mathbf{d} = d \sqrt{4 \pi/3} Y_p^1(\theta,\phi)$ ($p=0, \pm 1$) is a component of the dipole operator $\mathbf{d}$ where $\mathbf{\hat{e}_0} = \mathbf{\hat{z}},\mathbf{\hat{e}_{\pm 1}} = \mp (\mathbf{\hat{x}}\pm i \mathbf{\hat{y}})/\sqrt{2}$, $d$ is the permanent dipole moment of the molecule, and $Y_p^1$ are spherical harmonics.

\begin{figure}
    \centering
        \includegraphics[scale=.299]{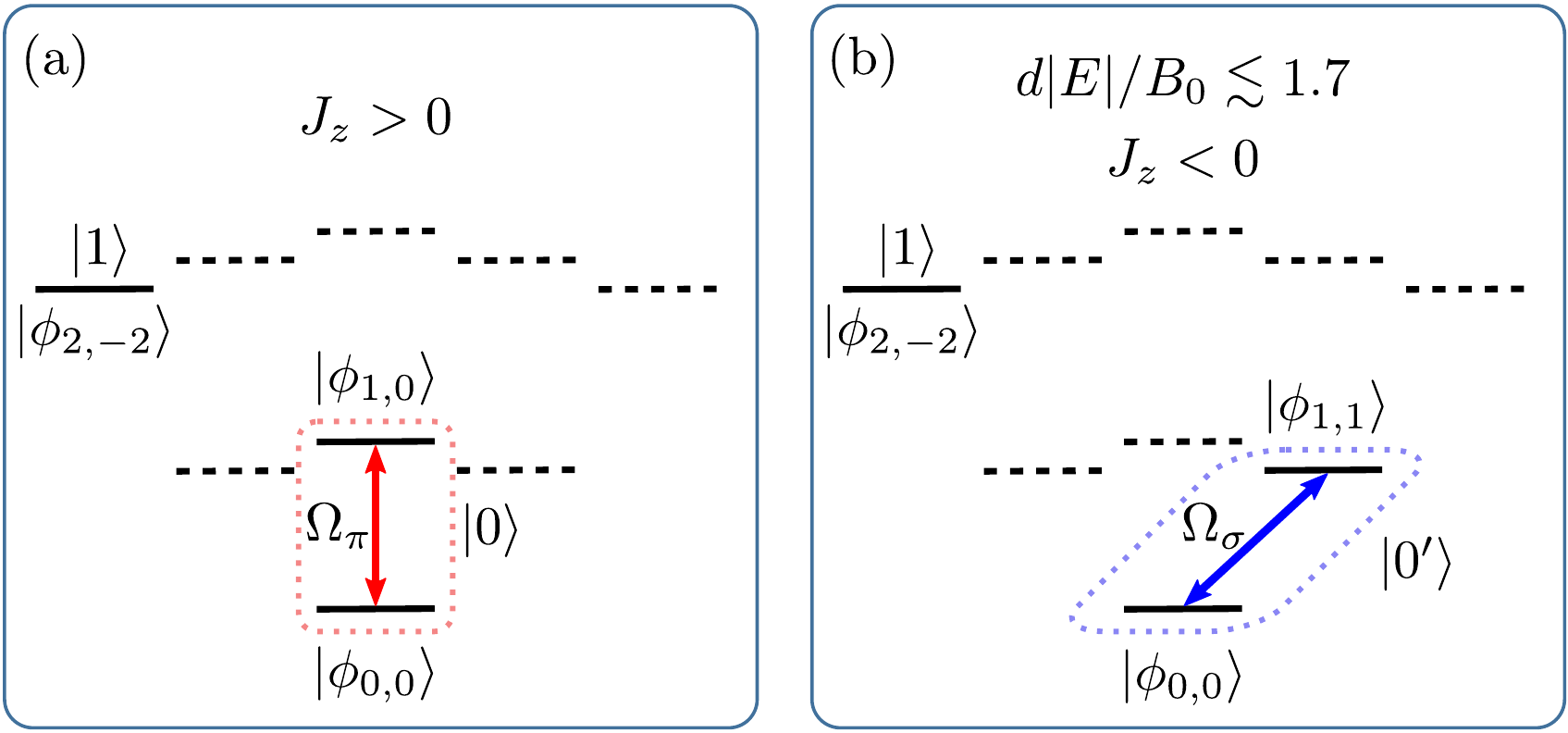}
    \caption{Schemes for realizing dipole-dipole Ising interactions using polar molecules, where $J_z$ denotes the strength of the Ising interactions. The polarization of the drive determines whether (a) positive $J_z$ or (b) negative $J_z$ is realized. Each approach works both in the presence or absence of a DC electric field, although the interactions realized in (b) exhibit positive $J_z > 0$ if $d|E|/B_0 \gtrsim 1.7$.
    }
    \label{fig:polarschemes}
\end{figure}

We will encode the computational states in the states $|N,M\rangle$, which are eigenstates of the angular momentum operator $\mathbf{N}^2$ and its projection $N_z$ on the quantization axis in the electronic and vibrational ground state of the molecule.
Defining $|\phi_{N,M}\rangle$ as the states which are adiabatically connected (as we turn on $E$) to $|N,M\rangle$,  we consider in scheme (a) (see Fig.~\ref{fig:polarschemes}) dressing the $|\phi_{0,0}\rangle$ and $|\phi_{1,0}\rangle$ states to engineer the $|0\rangle$ state and use $|\phi_{2,-2}\rangle$ as the $|1\rangle$ state.
The $|\phi_{0,0}\rangle$ and $|\phi_{1,0}\rangle$ states will be dressed via a resonant microwave drive with $\pi$ polarization and Rabi frequency $\Omega_\pi$, defining $|0\rangle = (|\phi_{0,0}\rangle + |\phi_{1,0}\rangle)/\sqrt{2}$. Although our method can be generalized to off-resonant drives, we will focus on a resonant drive since it provides the strongest interactions in the absence of an electric field, but we discuss the case of off-resonant drive in Appendix \ref{app:polar}.
Since the electric field does not generate a transition dipole moment between $|0\rangle$ and $|1\rangle$, no flip-flop interactions are introduced.
We note that, although our formalism will consider a nonzero electric field, our method works without issue in the absence of an electric field as well. Additionally, throughout, we will assume that the drive strength is always sufficiently strong that the hyperfine structure of the states being dressed can be neglected.

The relevant dipole moments of interest are $\mu_0 = \langle \phi_{0,0}| d_0 | \phi_{0,0}\rangle$, $\mu_1 = \langle \phi_{1,0}| d_0 | \phi_{1,0}\rangle$, $\mu_2 = \langle \phi_{2,-2}| d_0 | \phi_{2,-2}\rangle$, and $\mu_{01} = \langle \phi_{0,0}| d_0 | \phi_{1,0}\rangle$. In units of $(4 \pi \varepsilon_0)^{-1}$, the corresponding dipole-dipole interactions between $|0\rangle$ and $|1\rangle$ are (see Appendix \ref{app:polar})
\begin{subequations}
\begin{equation}
    \begin{split}
    H_\text{int} = \frac{1}{2}\sum_{i \neq j} \frac{1}{r_{ij}^3}& \left[(V_0+V_1-2 V_{01}) Z_i Z_j  \right. + \\
    & \left. (V_0-V_1)(Z_i + Z_j) + (V_0+V_1-2 V_{01}) \right]
    \end{split}
\end{equation}
\begin{gather}
    V_0 = \frac{(\mu_0 + \mu_1)^2 + 2 \mu_{01}^2 }{16}, \quad V_1 = \frac{\mu_2^2}{4}, \\
    V_{01} = \frac{(\mu_0 + \mu_1) \mu_2}{8},
 \end{gather}
\end{subequations}
which has an Ising interaction proportional to $V_0 + V_1 - 2V_{01} > 0$ for all $d |E|/B_0$. There are two considerations to avoid leakage to other states.
First, there are the resonant flip-flop interactions present in the absence of an electric field. As long as $\Omega_\pi \gg H_\text{dd}$, these will be made off-resonant through the dressed energies. Second, when the electric field is turned on, additional potentially near-resonant interactions emerge due to the induced dipole moments, such as flip-flop interactions between the $|\phi_{1,0}\rangle$, $|\phi_{1,1}\rangle$ states. Here, the dressed energies will typically keep these interactions off-resonant, although care has to be taken when $d |E| \approx |\Omega_\pi|$, as the energy shifts due to the electric field could induce resonances in combination with the dressed energies.

In scheme (b), see Fig.~\ref{fig:polarschemes}, to realize negative interactions, we instead consider dressing the $|\phi_{0,0}\rangle$ and $|\phi_{1,1}\rangle$ states to engineer the $|0'\rangle$ state and continue to use $|\phi_{2,-2} \rangle$ as the $|1\rangle$ state. The $|\phi_{0,0}\rangle$ and $|\phi_{1,1}\rangle$ states will be dressed via a resonant microwave drive with $\sigma_+$ polarization and Rabi frequency $\Omega_\sigma$, defining $|0'\rangle = (|\phi_{0,0}\rangle + |\phi_{1,1}\rangle )/\sqrt{2}$. Again, we will focus on the case of a resonant drive, which realizes the strongest interactions in the absence of an electric field.  Like with scheme (a), although our formalism assumes a nonzero electric field, scheme (b) also works in the absence of an electric field. However, we note that at sufficiently strong electric fields, the sign of the interactions can become positive.

The additional relevant dipole moments of interest are $\mu_1' = \langle \phi_{1,1}|d_0 |\phi_{1,1}\rangle$ and $\mu_{0,1}' = \langle\phi_{0,0}|d_-| \phi_{1,1}\rangle$. Once again, in units of $(4 \pi \varepsilon_0)^{-1}$, the corresponding dipole-dipole interactions between $|0'\rangle$ and $|1\rangle$ are
\begin{subequations}
\begin{equation}
    \begin{split}
    H_\text{int} = \frac{1}{2}\sum_{i \neq j} \frac{1}{r_{ij}^3}& \left[(V_0'+V_1-2 V_{01}') Z_i Z_j  \right. + \\
    & \left. (V_0'-V_1)(Z_i + Z_j) + (V_0'+V_1-2 V_{01}') \right]
    \end{split}
\end{equation}
\begin{gather}
    V_0' = \frac{(\mu_0 + \mu_1')^2 - \mu_{01}'^2 }{16}, \quad V_1 = \frac{\mu_2^2}{4}, \\
    V_{01}' = \frac{(\mu_0+\mu_1') \mu_2}{8},
\end{gather}
\end{subequations}
which has an Ising interaction proportional to $V_0' + V_1 - 2 V_{01}'$, which is less than 0 for for $d|E|/B_0 \lesssim 1.7$.
This value is modified by off-resonant drives, increasing or decreasing its value depending on the relative superposition of the two states being dressed (see Appendix \ref{app:polar}).
Like for scheme (a), leakage to other states can be avoided with sufficiently strong $\Omega_\sigma$ compared to $H_\text{dd}$ and by ensuring the energy shifts due to the electric field do not induce any new resonances.

Thus, by switching between the two schemes and simultaneously transferring the $|0\rangle$ population to the $|0'\rangle$ state, it is possible to implement the state-transfer protocols, although it is important to avoid unwanted interactions when switching between the schemes and states. Provided the relevant time-scales are fast enough relative to the interactions, one can transfer the states by adiabatically turning $\Omega_\pi$ off while adiabatically turning $\Omega_\sigma$ on in a STIRAP-like process, which simultaneously changes between the two schemes and changes the sign of the interactions as long as $d|E|/B_0 \lesssim 1.7$.

At the end of the state-transfer protocol, it is important to shelve the qubits in non-interacting states. A natural approach is to adiabatically turn off the DC electric and microwave dressing fields. By adiabatically detuning the dressing field simultaneously, this ensures that the $|0\rangle$ or $|0'\rangle$ state will be transferred to either of the two states which are dressed, while $|1\rangle$ becomes $|2,-2\rangle$.
Alternatively, one can keep the fields on by storing the qubits in pairs of states which possess the same electric dipole moment, and the resulting interactions will only generate a global phase on each individual qubit. In this case, all qubit states must be transferred simultaneously.
For example, this might be achieved by storing the qubit in $|\phi_{N, \pm M} \rangle$ pairs of states.
Nuclear spin states provide another natural means of realizing this while utilizing a single $|\phi_{N,M}\rangle$ state. Although we refer to these as non-interacting states, in reality they are states which exhibit state-insensitive, identity interactions both with each other and with the states $|0\rangle$ and $|1\rangle$ (or $|0'\rangle$ and $|1\rangle$) used in the protocol, although the strength of the identity interactions may be different.

Non-interacting states can in principle be prepared for use during the protocols. For example, in the absence of an electric field, any states without a transition dipole moment with the dressed states can be utilized as long as the two chosen states have identity interaction with each other, like nuclear spin states as above. This is more difficult in the presence of an electric field, as the permanent dipole moments will typically cause interactions with the $|0\rangle, |0'\rangle,$ and $|1\rangle$ states. By choosing parameters such that the induced dipole moments of $|0\rangle$ and $\ket{1}$ (or $|0'\rangle$ and $\ket{1}$) are the same, while the two non-interacting states have the same induced dipole moment and identity interactions with each other, the interactions with the non-interacting qubits become state-insensitive, leading only to a global phase. When the induced dipole moment of $|0\rangle$ and $\ket{1}$ are not the same, the interactions with the non-interacting states lead to $Z$ rotations in $|0\rangle$ and $\ket{1}$, which must be corrected (likewise for $|0'\rangle$ and $\ket{1}$).

Now, we determine when these single qubit corrections can be avoided for the two schemes. When $E \neq 0$, for scheme (a), this requires $\mu_2$ be between $\mu_0$ and $\mu_1$ while for scheme (b), this requires $\mu_2$ be between $\mu_0$ and $\mu_1'$. While this is possible for scheme (a) with the proper choice of drive and detuning (changing the relative contributions of $\mu_0, \mu_1$ to the induced dipole moment of $|0 \rangle$), it is not possible for scheme (b). However, if $E = 0$, this is automatically satisfied in both cases. Hence there are no single qubit corrections needed when there is no electric field.

\subsubsection*{Asymmetric blockade}
\label{sec:asymmetric}

In this section, we discuss the realization of microwave-dressed asymmetric blockade for polar molecules, which was developed for Rydberg atoms in Ref.~\cite{Young2021}. This provides a means of realizing the desired control-target interactions without needing to engineer both signs of interactions to eliminate the control-control and target-target interactions. Here, we discuss how this approach can be extended to polar molecules. Aside from a few subtleties which we discuss further below, the original analysis for the Rydberg atoms is largely the same, so we leave several of these details to Appendix \ref{app:polar}.

In contrast to the microwave-dressing approaches above, we consider dressing three states with two different drives, one which is $\pi$-polarized and one which is $\sigma_\pm$-polarized, with no electric field. We denote two of the three resulting dressed states $|c\rangle, |t\rangle$, which are used as controls or targets, respectively. Through proper tuning of the drives, these states exhibit asymmetric blockade: no control-control or target-target interactions but strong diagonal dipolar control-target interactions. This is achieved by counterbalancing the respective dipolar interactions associated with the two driven transitions, which differ in their overall sign due to the polarization of the transition, as utilized in the previous dressing schemes to realize both signs of interactions. The combination here eliminates the necessity to realize both positive and negative interactions because the unwanted interactions between controls and targets are already zero.
Thus if we use the $|c\rangle$ ($|t\rangle$) state to encode the $|0\rangle$ state for the control (target) molecules and an additional non-interacting state (e.g., from $N=4$) to encode $|1\rangle$ for both, we realize the interactions
\begin{equation}
    H_\text{int} = \sum_{i \in \{C\}, j \in \{T\}} \frac{C_3}{4 r_{ij}^3} (1-Z_i) (1-Z_j),
\end{equation}
where $C_3$ is the
coefficient describing the strength of the dipolar $c$-$t$ interactions and $\{C\},\{T\}$ define the sets of control and target atoms, respectively.

\begin{figure}
    \centering
    \includegraphics[scale=.3]{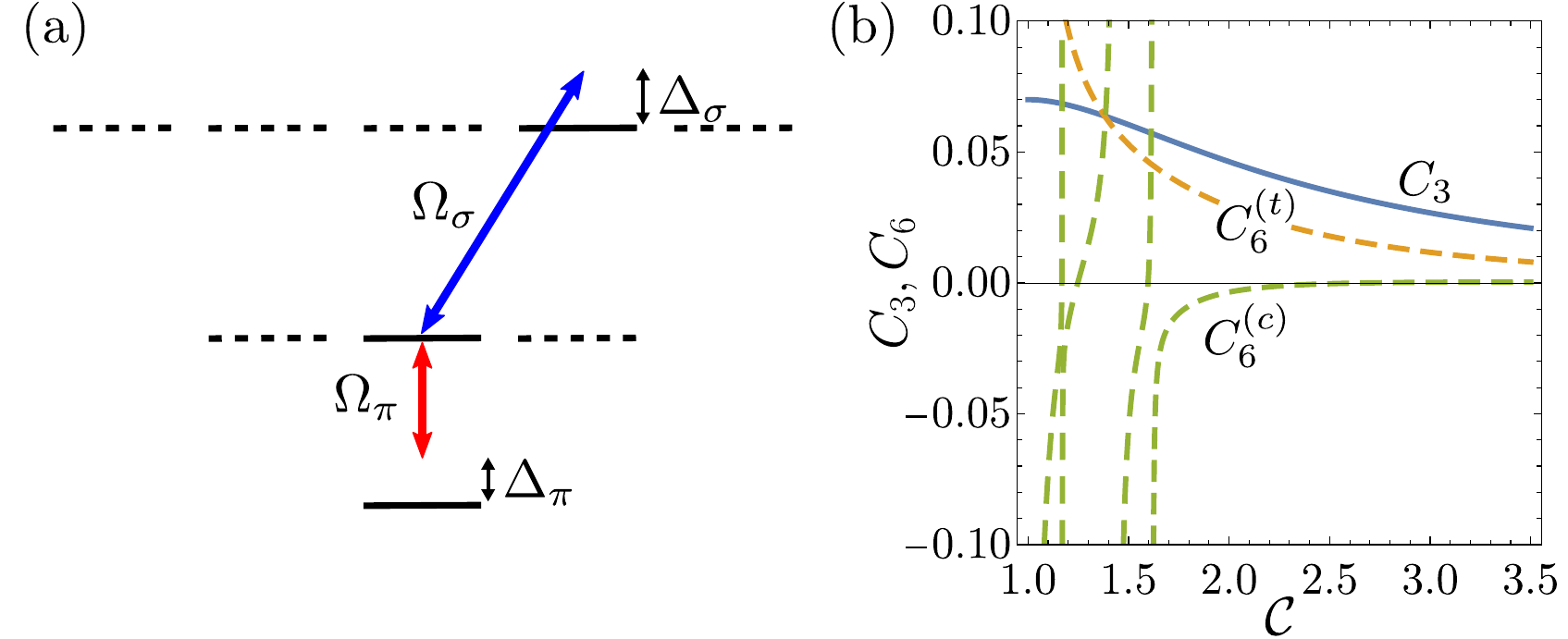}
    \caption{(a) Example of dressing scheme used to realize asymmetric blockade in polar molecules. The dressed $|0\rangle, |1\rangle$ states are composed of the molecular states denoted by solid lines. The remaining dashed levels contribute to the vdW interactions. (b) Dipolar c-t interaction coefficient ($C_3$, solid blue line), vdW c-c interaction coefficient ($C_6^{(c)}$, dashed green line), and vdW t-t interaction coefficent ($C_6^{(t)}$, dashed orange line) as functions of $\mathcal{C}$. Values of $C_3$ are in units of $d^2/(4 \pi \varepsilon_0)$ and $C_6$ in units of $d^4/[\Omega_\pi(4 \pi \varepsilon_0)^2]$.}
    \label{fig:asymm}
\end{figure}

Here, we consider the particular example of Fig.~\ref{fig:asymm}(a), dressing states $|0, 0\rangle, |1,0\rangle,$ and $|2,1\rangle$ with the rotating-frame Hamiltonian
\begin{equation}
\begin{aligned}
    H_\text{mw} = & ~ \Omega_\pi (|0,0 \rangle \langle 1,0|+|1,0 \rangle \langle 0,0|)~+ \\
    & ~ \Omega_\sigma (|2,1 \rangle \langle 1,0|+|1,0 \rangle \langle 2,1|) +~ \\
    &~ - \Delta_\pi |0,0 \rangle \langle 0,0| - \Delta_\sigma |2,1 \rangle \langle 2,1|,
\end{aligned}
\end{equation}
where $\Omega, \Delta$ denote the Rabi frequencies and detunings of the corresponding drives. The resulting dressed states take the form
\begin{subequations}
\begin{equation}
    |c\rangle \propto c_\pi |0,0 \rangle + |1,0 \rangle + c_\sigma |2,1 \rangle,
\end{equation}
\begin{equation}
    |t\rangle \propto t_\pi |0,0 \rangle + |1,0 \rangle + t_\sigma |2,1 \rangle,
\end{equation}
\end{subequations}
in addition to a third dressed state which we will not use.
For this choice of states, the maximal dipolar interaction coefficient between $|c\rangle$ and $|t\rangle$ is $C_3 = 0.07 \frac{d^2}{4 \pi \epsilon_0 r^3}$, which has an associated $c_\pi^{\text{max}}$ coefficient  that uniquely defines the other coefficients. However, $|c\rangle, |t\rangle$ are degenerate in the rotating frame of the drives when this maximal interaction is realized, which can lead to additional unwanted dipolar interactions (e.g., $|cc \rangle \langle tt|$). By working with $c_\pi$ close to $c_\pi^\text{max}$ but not equal to it, which we quantify via $\mathcal{C} \equiv  c_\pi/c_\pi^{\text{max}}$, this degeneracy can be lifted while still realizing asymmetric blockade with a comparable interaction strength, see Fig.~\ref{fig:asymm}(b).

An important consideration is the emergence of $1/r^6$ vdW $|c c \rangle \langle c c|, |t t \rangle \langle t t|$ interactions with coefficients $C_6^{(c)}, C_6^{(t)}$, respectively, which can lead to errors. For Rydberg atoms, there were two degrees of freedom that could be utilized to tune the vdW interactions: $\mathcal{C}$ and the overall scale of $H_\text{mw}$. The former modifies the dressed states while the latter modifies the dressed state energies. However, for polar molecules the overall scale of $H_\text{mw}$ only determines the overall scale of the resulting vdW interactions. This is because unlike Rydberg atoms which have many internal Rydberg states contributing to the vdW interactions, for polar molecules the transition energies involved differ on the order of GHz, several orders of magnitude larger than feasible dipolar interactions and microwave dressing fields. Hence, the only contributions to the vdW interactions arise from the typically resonant dipole-dipole interactions (in the absence of external fields) that are made off-resonant by the dressing field, which thus defines the overall scale of the vdW interactions. In the example we consider here, these involve all of the $N=0,1,2$ states.

In Fig.~\ref{fig:asymm}(b), we plot both the dipolar coefficient $C_3$ and the vdW coefficients $C_6^{(c/t)}$ as a function of $\mathcal{C}$ in units of $d^2/(4 \pi \varepsilon_0)$ and $d^4/[\Omega_\pi (4 \pi \varepsilon_0)^2]$, respectively. First, we note that there are several divergences in the $C_6$ coefficients. These are a consequence of pair states composed of $|0\rangle$ or $|1\rangle$ becoming degenerate with other dressed pair states. Since this leads to leakage outside of the computational states, these parameters must be avoided. Additionally, we see that, for the example considered, while one of the $C_6$ passes through zero for certain values of $\mathcal{C}$, the other does not. Nevertheless, we remark that the overall scale that relates the strength of the vdW interactions to dipolar interactions is given by $d^2/(\Omega_\pi 4 \pi \varepsilon_0 r^3)$, i.e., the ratio of the dipolar interactions to the dressing fields. As before, our dressing scheme already requires $\Omega_\text{mw} \gg H_\text{dd}$, so this is already small to begin with, but we wish to minimize it as much as possible. If we assume $\Omega_\pi$ to be on the order of MHz and $d^2/(4 \pi \varepsilon_0 r^3)$ to be on the order of kHz, this ensures that any vdW interactions are at least 1000 times smaller. Furthermore, we remark that, if an additional dressing field were included, this would likely provide sufficient tunability to fully eliminate both vdW interactions, but we leave this direction to future work.

\subsection{Rydberg atoms}
\label{subsec:rydbergdd}

\begin{figure}
    \centering
    \includegraphics[scale=.31]{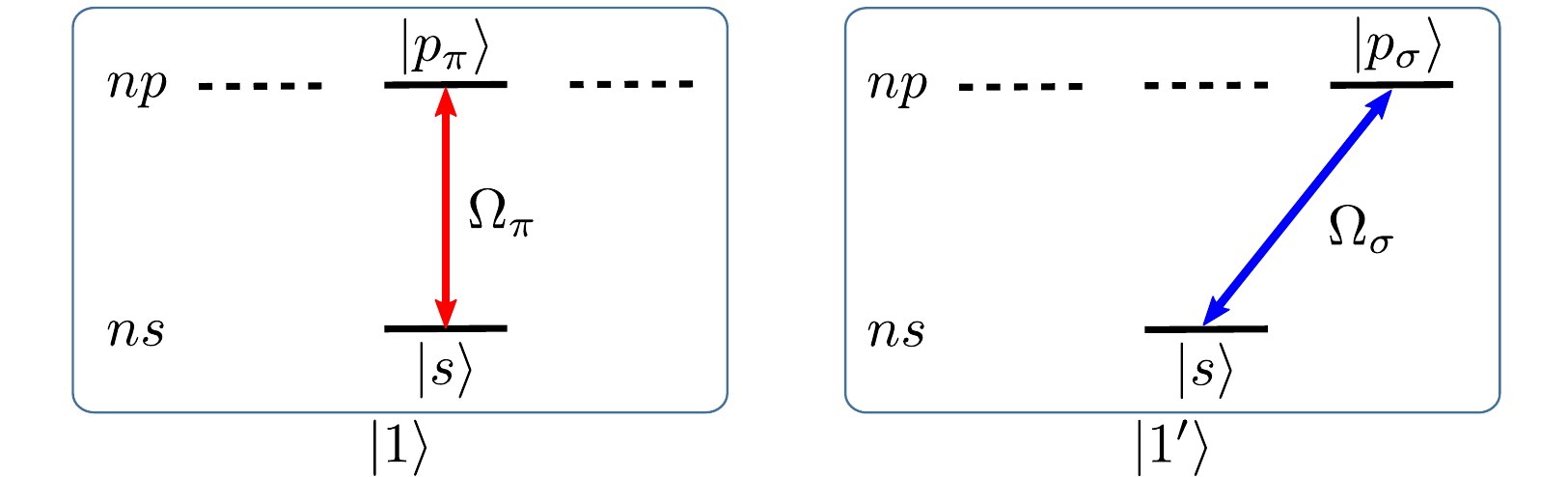}
    \caption{Dressing scheme for realizing (left) $H_\text{int}$ and (right) $H_\text{int}'$ via microwave dressing of Rydberg states, with $|1\rangle = (|s\rangle + |p_\pi\rangle)/\sqrt{2}$ and $|1'\rangle = (|s\rangle + |p_\sigma\rangle)/\sqrt{2}$. In each case, we require $\Omega_{0/+} \gg H_{\text{dd}}$ so that we may work in the microwave-dressed basis and make additional flip-flop interactions (e.g. those between $\ket{s}$ and the dashed-line states) off-resonant.}
    \label{fig:rydlevels}
\end{figure}

In this section, we consider the implementation of the state-transfer protocol using Rydberg atoms, which will share several similarities with the implementation in polar molecules. In order to do so, as shown in Fig.~\ref{fig:rydlevels},  we will consider encoding $|0\rangle, |1\rangle$ in linear combinations of four different states $|g\rangle$, $|s\rangle$, $|p_\pi\rangle$, and $|p_\sigma\rangle$, corresponding to a ground state, a Rydberg $s$ state, and two Rydberg $p$ states, respectively. In the case of the latter two states, $0/+$ correspond to the polarization of the drive needed to couple the $s$ state to the corresponding $p$ state. Additionally, we will assume that the $p$ states have similar values of the principal quantum number $n$ compared to the $s$ state, leading to strong dipole-dipole interactions between $s$ states and $p$ states. While Ref.~\cite{Eldredge2017} also realized the desired form of interactions for Rydberg atoms via a weak electric field, the scheme discussed here allows for much stronger interactions, providing faster state transfer with fewer errors due to dissipation.

In order to realize $H_\text{int}$, we will consider dressing the $|s\rangle$ state with the $|p_\pi\rangle$ state with a strong resonant microwave field. In the absence of any external fields, these two states interact according to the dipole-dipole interaction
\begin{equation}
    H_{\text{dd}} = \sum_{i\neq j} \mu_0^2 \frac{1- 3 \cos^2 \theta_{ij}}{r_{ij}^3} |s_i p_{j,\pi}\rangle \langle p_{i,\pi} s_j|,
\end{equation}
where $r_{ij}$ is the distance between atoms $i$ and $j$, $\theta_{ij}$ is the angle the displacement vector makes with the quantization axis, and $\mu_0 = \langle p_\pi| d_0| s \rangle$ is the corresponding transition dipole moment. Like for polar molecules, we have neglected additional resonant interaction terms involving other states because the introduction of a strong microwave field will cause these to become strongly off-resonant due to the resulting energy shifts. Additionally, since we are considering dipolar state transfer only in $d=2$, we shall consider $\theta_{ij} = \pi/2$ throughout.

Using the strong resonant microwave field, we may encode $|0\rangle = |g\rangle$ and $|1 \rangle = (|s\rangle + |p_\pi\rangle)/\sqrt{2}$. In the limit of a strong microwave Rabi frequency $\Omega_{\text{mw}} \gg H_{\text{dd}}$, we may express the dipole-dipole interactions in the $|0\rangle, |1\rangle$ basis as
\begin{equation}
    H_\text{int} = \frac{1}{2}\sum_{i \neq j} \frac{\mu_0^2}{8}  \frac{1}{r_{ij}^3} (1- Z_i ) (1-Z_j ).
\end{equation}

In order to realize $H'_{\text{int}} \equiv -a H_\text{int}$, we will use the same approach as for $H_\text{int}$. However, rather than dressing the $|s\rangle$ state with $|p_\pi\rangle$, we instead resonantly dress it with $|p_\sigma\rangle$. In this case, the corresponding dipole-dipole interaction is
\begin{equation}
    H_{\text{dd}}' = \sum_{i \neq j} -\frac{\mu_+^2}{2} \frac{1}{r_{ij}^3} |s_i p_{j,\sigma}\rangle \langle p_{i,\sigma} s_j|,
\end{equation}
where $\mu_+ = \langle p_\sigma |d_+ |s\rangle$ and we neglect the other terms due to the energy shifts caused by the strong microwave field once again. Note that, aside from a constant factor, the dipole-dipole interaction here only differs by a sign from $H_{\text{dd}}$. As such, if we now encode $|1'\rangle = (|s\rangle + |p_\sigma\rangle)/\sqrt{2}$ and again take the limit of a strong microwave Rabi frequency $\Omega_{\text{mw}} \gg H_{\text{dd}}$, the resulting interactions are
\begin{equation}
    H_\text{int}' = -\frac{1}{2}\sum_{i \neq j} \frac{\mu_+^2}{16} \frac{1-3 \cos^2 \theta}{r_{ij}^3} (1-Z_i) (1-Z_j),
\end{equation}
which differs from $H_\text{int}$ by a negative constant factor.

Like for polar molecules, by  transferring the $|1\rangle$ population to the $|1'\rangle$ state while also changing the polarization of the microwave field, it is possible to implement the state-transfer protocols via Rydberg atoms. Note that when the polarization of the drive is changed, one must ensure that the qubit state is temporarily not encoded in $|s\rangle, |p_\pi\rangle,$ or $|p_\sigma\rangle$. Alternatively, provided the relevant time-scales are fast enough relative to the interactions, one can transfer the states by adiabatically turning $\Omega_\pi$ off while adiabatically turning $\Omega_\sigma$ on, which simultaneously changes between the two schemes and changes the sign of the interactions. Additionally, to avoid leakage out of the computational basis, like with the polar molecules we utilize the fact that $\Omega \gg H_\text{dd}$, so the dressed energies will make these leakage processes off-resonant.

As discussed in the polar molecules section, it is also possible to realize the desired control-target interactions without the spin-echo approach by engineering asymmetric blockade interactions. This is achieved by turning on both drives $\Omega_{\pi/\sigma}$ with detunings $\Delta_{\pi/\sigma}$ and using different principal quantum numbers $n$ for $|p_{\pi/\sigma} \rangle$. With a proper choice of parameters, two of the resulting dressed states $|c\rangle, |t\rangle$ interact with each other but not themselves, realizing asymmetric blockade. Thus by forming the control qubits with $|c\rangle$ and the target qubits with $|t\rangle$ as the logical $|1\rangle$ states and the ground state as $|0\rangle$ for both, the desired pure control-target interactions can be realized. The full details of engineering this form of interactions for Rydberg atoms be found in Ref.~\cite{Young2021}.

\subsubsection*{Rydberg vdW interactions in 3D}
\label{subsec:rydbergvdw}
We briefly discuss how to use $1/r^6$ vdW interactions for state transfer in 3D. While this will not be as fast as using $1/r^3$ dipole-dipole interactions ($t \sim e^{3 \sqrt{3 \log L}}$ for vdW interactions with the Tran protocol vs.~$t \sim \polylog(L)$ for the Tran and Eldredge protocols with dipolar interactions), some aspects are less complicated since the vdW interactions will have no significant angular dependence. Moreover, it provides a natural means of experimentally demonstrating that vdW interactions are long-range in three dimensions.

One option to realize the desired sign change in the vdW interactions is by inducing a F\"{o}rster resonance for an $s$ state via an external magnetic field. Provided one keeps the atoms sufficiently far apart outside of the dipole-dipole interaction regime, where the interaction causes pairs of $s$ states to hybridize with pairs of $p$ states, the sign of $C_6$ will be opposite on either side of the F\"{o}rster resonance \cite{Saffman2010}.

Alternatively, microwave drive approaches can be used here as well. In this case, we would use an asymmetric blockade approach as in Ref.~\cite{Young2021} applied to vdW interactions. Ref.~\cite{Shi2017} discusses how a microwave-dressed superposition of Rydberg $s$ states can experience no vdW interactions. By engineering two such states with two pairs of Rydberg $s$ states, only the interactions between the two states remains. By separating these two pairs of $s$ states from each other in $n$, the flip-flop interactions will fall off rapidly compared to the diagonal vdW interactions as in Ref.~\cite{Belyansky2019}, leaving only the desired Ising interactions.

\begin{figure*}
    \centering
    \includegraphics[scale=.29]{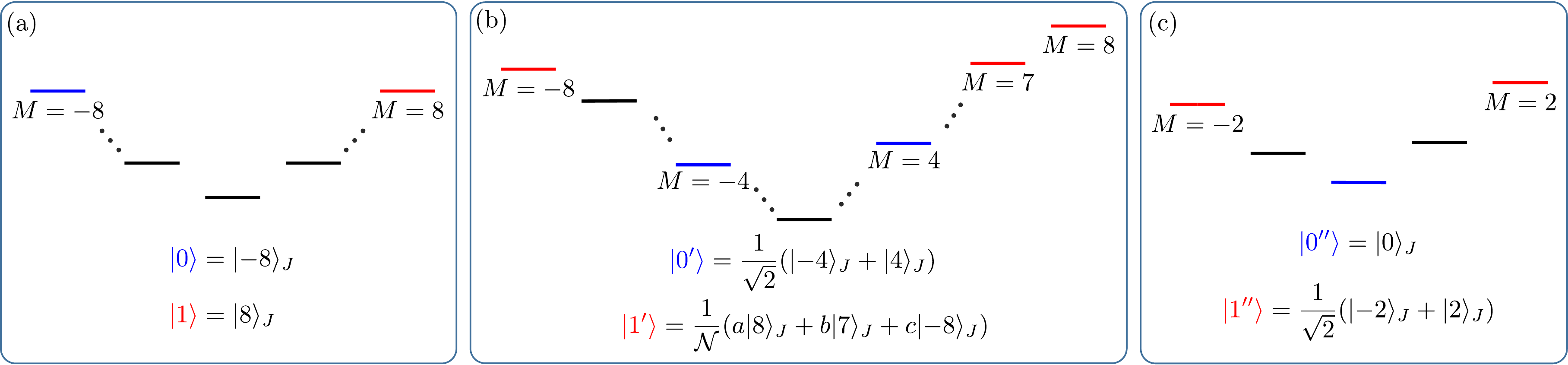}
    \caption{Level structure and dressing schemes for bosonic dysprosium atoms. We show only a subset of the full $J=8$ manifold. In all three schemes, a far-detuned optical field is included, while a magnetic field is also applied in (b) and (c). (a) Dressing scheme for realizing positive interactions. (b) Dressing scheme for realizing negative interactions. (c) Dressing scheme for realizing non-interacting states. }
    \label{fig:Dy_levels}
\end{figure*}

\subsection{Magnetic atoms (dysprosium)}
\label{sec:dy}

In this subsection, we consider the implementation of the state-transfer protocol in a two-dimensional lattice of ultracold dysprosium atoms.
For concreteness, we consider bosonic isotopes of dysprosium (such as $^{162}$Dy, $^{164}$Dy), so that the nuclear spin is zero.
We choose to encode the computational states $\ket{0},\ket{1}$ in the Zeeman manifold of the electronic ground state of the dysprosium.
The interactions between a pair of dysprosium atoms $i,j$ are given by the following magnetic dipole-dipole interactions---written in units of $\mu_0(g_J\mu_B)^2/4\pi$, where $g_J$ is the Land\'e $g$ factor, $\mu_B$ is the Bohr magneton, and $\mu_0$  is the vacuum magnetic permeability (not to be confused with the dipole transition moment used previously) \cite{Yao2015}:
\begin{equation}
	\label{eq:H_dipdip-mag}
 \begin{split}
	H_{ij} = \frac{1}{r_{ij}^3}\Big[(&1-3\cos^2\theta_{ij})[J_i^zJ_j^z-\frac{1}{4}(J_i^+J_j^-+J_i^-J_j^+)]\\
    &-\frac{3}{4}\sin^2\theta_{ij} [e^{-2i\phi_{ij}}J_i^+J_j^++\text{H.c}] \Big].
 \end{split}
\end{equation}
Here, $J=8$ is the electronic angular momentum of the ground state, and $\theta_{ij},\phi_{ij}$ are the polar and azimuthal angles of the displacement vector between sites $i$ and $j$ with respect to the quantization axis, which is set by a far-detuned optical field. This field additionally breaks the degeneracy of the Zeeman levels $\ket{M}_J, -8 \leq M \leq 8$, which can be utilized to limit leakage out of the computational basis. Since the corresponding energy shifts are quadratic in $M$, in some cases it might also be necessary utilize a linear energy shift in $M$ via a Zeeman shift, further breaking degeneracies.

In the following, we will describe example dressed states that can realize the desired forms of interactions. While we will not discuss precise dressing schemes to prepare these dressed states, the same principles will hold as for polar molecules and Rydberg atoms. Specifically, the dressing field must be stronger than the resulting dressed interactions. Additionally, it is important to ensure that the interactions do not lead to leakage out of the computational basis. While typically we expect the dressing to ensure any such processes will be off-resonant, we will discuss which leakage processes need to be considered.
As always, we restrict our focus to $\theta_{ij} = \pi/2$, which here allows us to neglect the $J_i^+ J_j^+$ terms in the Hamiltonian as well.

The simplest way of realizing the diagonal interactions that we require is to make use of the $J_i^zJ_j^z$ term in \cref{eq:H_dipdip-mag}, setting $\ket{0}\equiv\ket{-m}_J,\ket{1}\equiv\ket{m}_J$ with $\abs{m}\ge 2$, as shown in \cref{fig:Dy_levels}(a). The choice $m=8$ maximizes the interaction strength and yields the Hamiltonian
\begin{equation}
    H_\text{int} = \frac{1}{2} \sum_{i\neq j} \frac{64}{r_{ij}^3}Z_iZ_j
\end{equation} 
Flip-flop interactions can
be avoided by once again using an off-resonant optical field. Neither a magnetic field nor a dressing field is necessary in this case since we utilize the Zeeman levels directly.

\begin{figure*}
\centering
   \includegraphics[width=\textwidth]{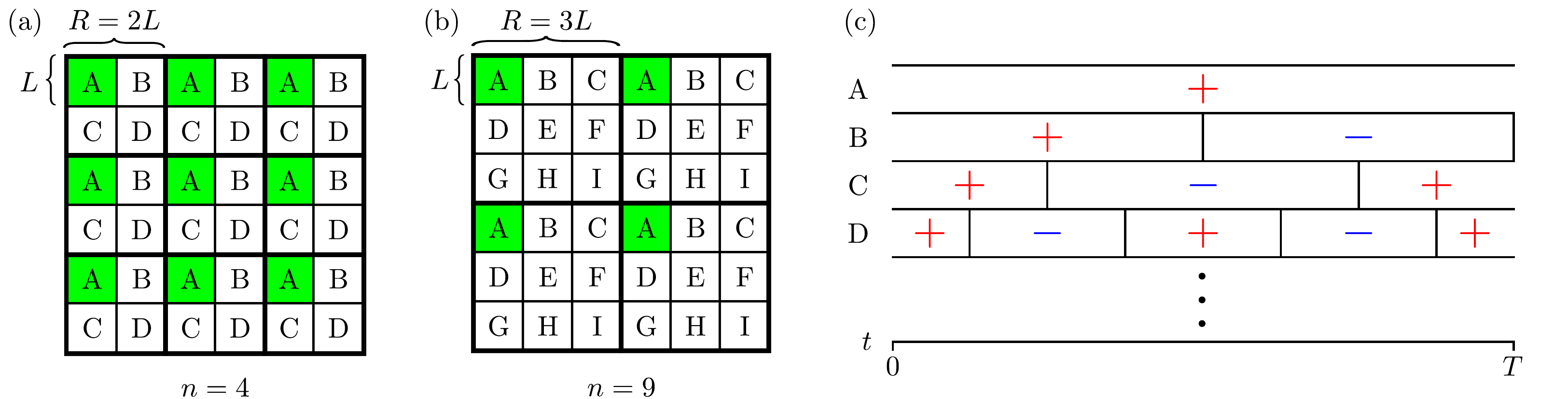}
    \caption{Scheme for implementing different colors in 2D. Each plaquette only interacts with other plaquettes of the same ``color,'' e.g. plaquettes labeled A (green) only talk to each other.
    For $n$ different colors, the distance between plaquettes of the same color scales as $R = \sqrt{n} L$, where $L$ is the size of the individual plaquette.
    In subfigures (a) and (b), we depict the Tran protocol generating a GHZ state with characteristic length $r = 6L$ using $n=4$ and $n=9$ colors respectively; the distances between plaquettes of the same color will be $R=2L$ and $R=3L$ respectively.
    (c) The spin-echo pulse sequence that cancels out interactions between plaquettes with different colors. Vertical lines between the $+$'s and $-$'s denote the spin-echo pulses. For $n$ colors, the protocol will in general require $2^{n-1}$ distinct pulses.}
		\label{fig:crosstalk_plaquettes}
\end{figure*}

In order to realize the reverse Hamiltonian, i.e the same Hamiltonian but with opposite sign, we will make use of the flip-flop interactions in \cref{eq:H_dipdip-mag}.
Using microwave and/or optical dressing, we can choose, for example, the following dressed states: $\ket{0'}\equiv(\ket{-4}_J+\ket{4}_J)/\sqrt{2}$, and $\ket{1'}\equiv\frac{1}{\mathcal{N}}(a\ket{8}_J+b\ket{7}_J+c\ket{-8}_J)$, where the $a,b,c$ amplitudes are tunable via the dressing parameters and $\mathcal{N}$ is a normalization constant [\cref{fig:Dy_levels}(b)].
This choice clearly yields $\bra{0'} J_z\ket{0'}=0$ and by properly tuning $a,b,c$ (namely, setting $\abs{b}^2=8(\abs{c}^2-\abs{a}^2)/7$) we can also ensure that $\bra{1'} J_z\ket{1'}=0$. In this case, only the flip-flop interactions from \cref{eq:H_dipdip-mag} contribute, for which the only nonzero matrix element is $\bra{1'1'}H_{ij}\ket{1'1'}$. This realizes the new interaction Hamiltonian \begin{equation}
H'_\mathrm{int}= -\frac{1}{2}\sum_{i\neq j}\frac{2\abs{ab}^2}{r_{ij}^3\abs{\mathcal{N}}^4}(1-Z_i)(1-Z_j),
\end{equation} with the same $Z_iZ_j$ interactions as before, but with the opposite sign.

Finally, we describe how to realize non-interacting states. We cannot use the bare Zeeman levels, since they will always have non-trivial interactions, particularly due to the $J_i^zJ_j^z$ term in \cref{eq:H_dipdip-mag}. Thus, we propose to use the dressed states $\ket{0''}\equiv|0\rangle_J,\ket{1''}\equiv  (\ket{-2}_J+ \ket{2}_J)/\sqrt{2}$ (which can be realized using microwave or optical dressing) to encode the computational $\ket{0''},\ket{1''}$ states, as shown in \cref{fig:Dy_levels}(c). Due to the choice of states, both have $\langle J_z\rangle = 0$ and are not coupled via the flip-flop interactions. We could not have used $(\ket{-2}_J - \ket{2}_J)/\sqrt{2}$ as the $|0''\rangle$ state (which could be achieved with no dressing in the absence of a Zeeman shift) because the Ising interactions would induce resonant flip-flop interactions in the computational basis. Thus dressing is necessary to realize non-interacting states.

In schemes (b) and (c), it is important in to ensure that the bare $H_{ij}$ interactions do not induce leakage to states outside of the computational basis. For example in (b), the flip-flop interactions can take $|7\rangle_J|8\rangle_J \to |8\rangle_J |8\rangle_J$ or $|{-4}\rangle_J |4 \rangle_J \to |{-3}\rangle_J |5\rangle_J$, while the Ising interactions could take $|0'\rangle |0'\rangle \to (|{-4}\rangle_J - |4\rangle_J)(|{-4}\rangle_J - |4\rangle_J)/2$. While in (a) these are automatically off-resonant due to the far-detuned optical field or 0 due to the choice of states, similar processes exist for (c). Although these interactions may or may not be resonant in the undressed basis, this can change in the dressed basis. In light of this, it is important that any dressed states which have overlap with the bare Zeeman levels involved in these interaction have dressed energies which ensure that any potential leakage is off-resonant. In some cases, the dressing fields will be sufficient to ensure that they are off-resonant; however, in cases where this is not the case, for example due to the $M \to -M$ symmetry, the Zeeman shift (in addition to the far-detuned optical field) provides another degree of tunability which simultaneously removes any accidental resonances from the symmetry.

\section{Crosstalk interactions}
\label{sec:crosstalk}

One concern in the implementation of the protocol by Tran \etal \cite{Tran2021a} is that, in the process of simultaneously creating GHZ states on multiple registers of qubits, crosstalk interactions may occur between qubits in different clusters.
Specifically, while the spin-echo procedure is able to eliminate interactions within and between registers of control (target) qubits, the control qubits in one register may interact with target qubits in another register, leading to undesirable crosstalk interactions.
In this section, we analyze the extent of these crosstalk interactions and propose a few methods to address them.

Recall from \cref{sec:state-transfer-protocols} that, in the Tran protocol, in order to encode a qubit into a generalized GHZ state of characteristic length $r$, it is first necessary to create $m^d$ GHZ states (including one generalized GHZ state) with characteristic length $r_1 \equiv r/m$ in parallel, where $m$ is some $\alpha$-dependent factor.
In \cref{fig:crosstalk_plaquettes}, we illustrate the set-up using qubits arrayed in a 2D grid.
We take $L = r_1$ to be the characteristic length of one of the $m^2$ ($d=2$) plaquettes $C_1, C_2, \dots, C_{m^2}$---each containing $L^2$ qubits.
Then, crosstalk interactions may arise due to the $L^{4}$ pairwise couplings present between qubits in different plaquettes.

Using a variation of the spin-echo protocol with additional single-qubit $\pi/2$ pulses, it's possible to cancel the interactions between controls and targets in different registers as follows.
We label the $m^d$ registers with $n$ different ``colors'' and apply single-qubit pulses that cancel the crosstalk interactions between the qubits in registers with different colors.
The pulse sequence for the protocol is shown for $n=4$ in \cref{fig:crosstalk_plaquettes}(c).
The sequence of single-qubit pulses will lead to alternating signs of the interaction Hamiltonians and lead to cancellations between registers with different colors.
However, interactions will still occur between qubits in registers with the same color.
To estimate the crosstalk error incurred by these interactions, we perform the following analysis.

Consider in the worst case that all qubits in plaquettes with the same color will interact with each other during a single GHZ-creation step.
Then, multiply the total strength of this crosstalk interaction by the number of colors and the time required to create the GHZ state, and then sum this quantity over the total number of recursive steps required to implement the full protocol.
This leads to the following bound on the total crosstalk error [full details in \cref{app:crosstalk_analysis}]:
\begin{align}
	\label{eq:crosstalk}
	\eps_\text{tot} =
	\begin{cases}
		\O{\frac{r^{2d-\alpha}}{n^{\alpha/d-1}} \polylog(r)} & \text{if } d< \alpha < 2d,\\
 		\O{\frac{e^{{\gamma}\sqrt{\log r}}}{n} \polylog(r)}&\text{if } \alpha = 2d,  \\
		\O{\frac{1}{n^{\al/d-1}}\polylog(r)} & \text{if } 2d< \alpha\leq 2d+1.
	\end{cases}
\end{align}
As such, the number of colors required in order to achieve a fixed amount of crosstalk error [$\eps_\text{tot}=\O1$] scales as
\begin{align}
	\label{eq:colors}
	n =
	\begin{cases}
		\W{r^{d(2d-\alpha)/(\alpha-d)} \polylog(r)} & \text{if } d< \alpha < 2d,\\
		\W{e^{\gamma\sqrt{\log r}}\polylog(r)}& \text{if }\alpha = 2d, \text{and} \\
		\W{\log(r)^{d/(\al-d)}} & \text{if } 2d< \alpha\leq 2d+1,
	\end{cases}
\end{align}
where the $\Omega(\cdot)$ denotes an asymptotic lower bound.
Intuitively, for smaller $\al$, the interactions are stronger, so more colors are required to cancel out the error.
In particular, for $\al \le 2d$, the number of colors required to achieve a fixed crosstalk error scales super-logarithmically in the state-transfer distance.

In order to cancel the errors using the spin-echo procedure, the number of spin pulses required will scale exponentially in the number of colors.
Specifically, for color $i\in[n]$, a total of $2^{i-1}$ for $i \geq 3$ ($i=1,2$ require 0 and 2, respectively) single-qubit pulses will be required to provide the alternating signs required for the interactions [\cref{fig:crosstalk_plaquettes}(c)].
As such, for $\al < 2d$, the fact that $n$ scales at least polynomially in $r$ will lead to an exponential scaling of the number of pulses with $r$.
In the worst case, when performing a single step of the hybrid protocol, the number of colors will be at most equal to the total number of GHZ plaquettes, i.e.~$m^d$.
For $d=2$, we determined in \cref{sec:state-transfer-protocols} that the crossover point where the hybrid protocol surpasses the pure Eldredge protocol was for characteristic length $r=244$ and $m=5$.
For this value, the spin-echo procedure would in the worst case require $2^{m^2-1} \approx 1.6\times 10^7$ pulses.
As such, an approach based solely on spin-echo pulses would likely be inefficient at mitigating crosstalk errors.

Another approach to eliminating crosstalk between qubits in different registers would be to use multiple different level encodings. This approach would only require a number of encodings proportional to the number of colors required to achieve a fixed crosstalk error, leading to a resource cost that scales only polynomially in the state transfer distance $r$.

As an example, for Rydberg atoms, these different encodings could be realized through the use of multiple states with different principal quantum numbers to encode the corresponding $|1\rangle$ states for each color. Provided the principal quantum numbers are sufficiently different, the interactions between them are minimal, although the number of colors can become limited due to the large variation in the vdW coefficients $C_6$ as a function of the principal quantum number.

As another example, the asymmetric blockade approach, developed in Ref.~\cite{Young2021} and discussed in \cref{subsec:polar-molecules,subsec:rydbergdd}, could be generalized to eliminate the interactions between different colors through the use of additional dressing fields, whose number will scale at most polynomially in the number of colors (since the corresponding constraints on the interactions scale polynomially).
Used in conjunction with the spin-echo-based crosstalk mitigation strategy, these encoding-based approaches could potentially eliminate the crosstalk error using far fewer---indeed, exponentially so---spin-echo pulses.
Concretely, in the example discussed above with $m=5$, generating a pair of colors using state engineering would reduce the number of pulses to $2^{m^2/2-1} \approx 2900$---i.e. four orders of magnitude fewer than without state engineering. 
If five colors are realized, then the number of pulses drops to 16, which is well within the realm of experimental feasibility.

In summary, we've outlined two methods to eliminate the crosstalk interactions between control and target qubits in different registers incurred during the Tran protocol: the first by using additional spin-echo pulses, and the second by encoding qubits in different energy levels.
We observed that the latter method is able to mitigate crosstalk errors using a number of encodings that only scales polynomially in the state transfer distance.
We envision that a combination of the two methods will help experimentalists to manage the crosstalk interactions that occur throughout the Tran protocol in a thorough and efficient way.

\section{Conclusions and Outlook}
In conclusion, we've provided a roadmap for the experimental realization of fast state-transfer protocols in atomic and molecular systems.
The approaches we presented only require single-qubit control in addition to global power-law interactions, which can be realized using electric and magnetic dipole-dipole interactions and van der Waals interactions present in polar molecules, Rydberg atoms, and atoms with large magnetic moments.
For these types of interactions, the protocols are able to realize a superballistic---and, sometimes, even a superpolynomial---state transfer, as opposed to ballistic state transfer for finite-range interactions.

We also addressed certain issues that may arise throughout the course of implementing these protocols, namely the potential for crosstalk errors that could occur in the process of creating multiple many-body entangled states in parallel.
We presented a few methods that can 
mitigate these errors. 
One interesting further direction would be to look into whether the protocols themselves could be modified to not suffer from the crosstalk errors at all.

Finally, we note that these protocols can be quite sensitive to noise.
This is largely due to their use of the GHZ state as an intermediate entangled state, which is highly sensitive to single-qubit decoherence.
Other protocols that utilize more robust intermediate states could potentially provide higher noise tolerance.
For example, the protocol in Ref.~\cite{Hong2021} achieves state transfer with the same scaling in time as the Eldredge protocol, but uses an intermediate W state, which is known to have high robustness to single-qubit noise.
However, existing protocols using intermediate W states have typically required the use of flip-flop interactions \cite{Guo2020,Hong2021}, for which the spin-echo procedure for selectively implementing control-target interactions doesn't apply.
Additionally, since the protocol in Ref.~\cite{Hong2021} relies on uniform coupling, inhomogeneities in the interactions---whether from the power-law form or angular dependence---lead to fundamental errors in the state transfer that, unlike for the Eldredge and Tran protocols, cannot be eliminated through shelving sites in non-interacting states or through spin-echo techniques.
We leave the investigation of the experimental realization of such W-state-based protocols to future work.

\begin{acknowledgments}
We would like to thank Su-Kuan Chu, Minh Tran, Alicia Kollar, Andrew Lucas, Adam Ehrenberg, and Yifan Hong for useful discussions.
A.Y.G., R.B., P.B., and A.V.G.~were supported in part by the AFOSR MURI program, the DoE ASCR Quantum Testbed Pathfinder program (awards No.~DE-SC0019040 and No.~DE-SC0024220), AFOSR, NSF QLCI (award No.~OMA-2120757), DoE ASCR Accelerated Research in Quantum Computing program (award No.~DE-SC0020312), NSF STAQ program, and DARPA SAVaNT ADVENT.
Support is also acknowledged from the U.S.~Department of Energy, Office of Science, National Quantum Information Science Research Centers, Quantum Systems Accelerator.
J.T.Y.~was supported by the NIST NRC Research Postdoctoral Associateship Award, NIST, and the NWO Talent Programme (project number VI.Veni.222.312), which is (partly) financed by the Dutch Research Council (NWO).
\end{acknowledgments}

\appendix
\section{Review of optimal state-transfer protocol by Tran \etal}
\label{app:ost-protocol}
We first describe the setting of the problem and the main result of Ref.~\cite{Tran2021a} in this section.
For simplicity, we consider a $d$-dimensional hypercubic lattice $\Lambda$ and a two-level system located at every site of the lattice.
The protocol generalizes straightforwardly to all regular lattices.
Without loss of generality, we assume that the lattice spacing is one.
We consider a power-law interacting Hamiltonian
	$H(t) = \sum_{i,j\in \Lambda} h_{ij}(t), $
where $h_{ij}(t)$ is a Hamiltonian supported on sites $i,j$ such that, at all times $t$ and for all $i \neq j$, we have $\norm{h_{ij}}\leq 1/\dist({i,j})^\alpha$, where $\dist({i,j})$ is the distance between $i,j$, $\norm{\cdot}$ is the operator norm, and $\alpha\geq 0$ is a constant.
We use $\GHZab_S$ to denote the GHZ-like state over sites in $S\subseteq \Lambda$:
\begin{align}
	\GHZab_S
	\equiv a\ket{\bar0}_S + b\ket{\bar 1}_S,
\end{align}
where $\ket{\bar x}_S \equiv \bigotimes_{j\in S} \ket{x}_j$ $(x = 0,1)$ are product states over all sites in $S$ and $a,b$ are complex numbers such that $\abs{a}^2+\abs{b}^2 = 1$.
In particular, we use $\GHZ$ to denote the symmetric state $a = b = 1/\sqrt2$.

Given a $d$-dimensional hypercube $C\subseteq \Lambda$ of length $r\geq 1$, we consider the task of encoding a possibly unknown state $a\ket{0}+b\ket{1}$ of a site $c \in C$ into the GHZ-like state $\GHZab_C$ over $C$, assuming that all sites in $C$, except for $c$, are initially in the state $\ket{0}$.
Specifically, we construct a time-dependent, power-law interacting Hamiltonian $H(t)$ that generates $U(t) = \mathcal T \exp\left(-i\int_0^t ds H(s)\right)$ satisfying
\begin{align}
	&U(t)\ (a\ket{0}+b\ket{1})_{c} \ket{\bar 0}_{C\setminus c} = a \ket{\bar 0}_C + b \ket{\bar 1}_C\label{eq:UGHZ}
\end{align}
at time
\begin{align}
	t(r) \leq K_\alpha\times \begin{cases}
		\log^{\kappa_\alpha} r & \text{if } d< \alpha< 2d,\\
		e^{\gamma\sqrt{\log r}}& \text{if } \alpha = 2d, \\
		r^{\alpha-2d} & \text{if }2d<\alpha\leq 2d+1.
	\end{cases} \label{eq:Tbound}
\end{align}
Here, $\gamma = 3\sqrt{d}$, $\kappa_\al = \log(4)/\log(2d/\alpha)$, and $K_\alpha$ are constants independent of $t$ and $r$.
Additionally, by reversing the unitary in \cref{eq:UGHZ} to ``concentrate'' the information in $\GHZab$ onto a different site in $C$, we can transfer a quantum state from $c\in C$ to any other site $c'\in C$ in time $2t$.

The key idea of the protocol from \cite{Tran2021a} is to recursively build the GHZ-like state in a large hypercube from the GHZ-like states of smaller hypercubes.
For the base case, we note that hypercubes of finite lengths, i.e.\ $r\leq r_0$ for some fixed $r_0$, can always be generated in times that satisfy \cref{eq:Tbound} for some suitably large (but constant) prefactor $K_\alpha$.
Assuming that we can encode information into a GHZ-like state in hypercubes of length $r_1$ in time $t_1$ satisfying \cref{eq:Tbound}, the following subroutine encodes information into a GHZ-like state in an  arbitrary
hypercube $C$ of length $r = m r_1$ containing $c$---the site initially holding the state information $a,b$. Here $m$ is an $\alpha$-dependent number that we choose to be proportional to $r_1^{2d/\al-1}$ for $\al<2d$; to $\exp{\frac{\gamma}{2d}\sqrt{\log r_1}}$ for $\al=2d$; and to $r_1$ for $\al>2d$.

\textbf{Step 1:}
We divide the hypercube $C$ into $m^d$ smaller hypercubes $C_1,\dots,C_{m^d}$, each of length $r_1$.
Without loss of generality, let $C_1$ be the hypercube that contains $c$.
Let $V = r_1^d$ be the number of sites in each $C_j$.
In this step, we simultaneously encode $a,b$ into $\GHZab_{C_1}$ and prepare $\GHZ_{C_j}$ for all $j=2,\dots,m^d$, which, by our assumption, takes time
\begin{align}
	t_1  \leq K_\alpha\times\begin{cases}
		\log^{\kappa_\alpha} r_1 & \text{if } d< \alpha < 2d,\\
		e^{\gamma \sqrt{\log r_1}} &\text{if } \alpha = 2d,  \\
		r_1^{\alpha-2d} & \text{if }2d< \alpha\leq 2d+1.
	\end{cases} \label{eq:t1bound}
\end{align}
By the end of this step, the hypercube $C$ is in the state
\begin{align}
	(a\ket{\bar 0} + b\ket{\bar 1})_{C_1}\bigotimes_{j=2}^{m^d} \frac{\ket{\bar 0}_{C_j} + \ket{\bar 1}_{C_j}}{\sqrt{2}}.
\end{align}

\textbf{Step 2:}
Next, we apply the following Hamiltonian to the hypercube $C$:
\begin{align}
	H_2 = \frac{1}{(mr_1\sqrt{d})^\alpha} \sum_{j = 2}^{m^d} \sum_{\mu\in C_1} \sum_{\nu\in C_j} \ket{1}\bra{1}_\mu\otimes \ket{1}\bra{1}_\nu. \label{eq:cPhase}
\end{align}
This Hamiltonian effectively generates the so-called controlled-PHASE gate between the hypercubes, with $C_1$ being the control hypercube and $C_2,\dots, C_{m^d}$ being the target hypercubes.
We choose the interactions between qubits in \cref{eq:cPhase} to be identical for simplicity.
If the interactions were to vary between qubits, we would simply turn off the interaction between $C_1$ and $C_j$ once the total phase accumulated by $C_j$ reaches $\pi$
~\footnote{Because only the total accumulated phase matters in choosing the evolution time, we also expect the protocol to be robust against experimental errors such as uncertainties in the positions of individual particles:
If the position of each particle is known up to a precision $\epsilon\ll 1$, the total worst-case error in the accumulated phase scales as $t({r_1^{2d}}/{r^\alpha})\times (\epsilon/r)$, with $r_1$ being the length of each hypercubes and $r$ being the minimum distance between them. The result is a relative phase error proportional to $\epsilon/r$, which becomes smaller and smaller as the distance between the hypercubes increases.
Moreover, we expect the relative error to be even smaller in the commonly occurring situation when uncertainties in the positions are uncorrelated between different particles
}.
The prefactor $1/(mr_1\sqrt{d})^\alpha$ ensures that this Hamiltonian satisfies the condition of a power-law interacting Hamiltonian.
It is straightforward to verify that, under this evolution, the state of the hypercube $C$ rotates to
\begin{align}
	a
	\ket{\bar 0}_{C_1}\bigotimes_{j=2}^{m^d}\frac{\ket{\bar 0}_{C_j} + \ket{\bar 1}_{C_j}}{\sqrt 2}+
	b \ket{\bar 1}_{C_1}\bigotimes_{j=2}^{m^d}\frac{\ket{\bar 0}_{C_j} - \ket{\bar 1}_{C_j}}{\sqrt{2}}
\end{align}
after time $t_2 = \pi d^{\alpha/2}(mr_1)^\alpha/{V^2}$.

The role of power-law interactions in our protocol can be inferred from the value of $t_2$.
Intuitively, the speed of simultaneously entangling hypercube $C_1$ with hypercubes $C_2,\dots,C_{m^d}$ is enhanced by the $V^2 = r_1^{2d}$ couplings between the hypercubes.
However, the strength of each coupling, proportional to $1/(mr_1)^\alpha$, is suppressed by the maximum distance between the sites in $C_1$ and those in $C_2,\dots,C_{m^d}$.
With a small enough $\alpha$, the enhancement due to $V^2$ overcomes the suppression of power-law interactions, resulting in a small entanglement time $t_2$.
In particular, when $\alpha < 2d$, $t_2$ actually \emph{decreases} with $r_1$, implying that Step 2 would be faster in later iterations if we were to keep $m$ constant.

To obtain the desired state $\GHZab_C$, it remains to apply a Hadamard gate on the effective qubit $\{\ket{\bar 0}_{C_j},\ket{\bar 1}_{C_j}\}$ for $j = 2, \dots, m^d$. We do this in the following three steps by first concentrating the information stored in hypercube $C_j$ onto a single site $c_j \in C_j$  (Step 3), then applying a Hadamard gate on $c_j$ (Step 4), and then unfolding the information back onto the full hypercube $C_j$ (Step 5).

\textbf{Step 3:}
By our assumption, for each hypercube $C_j$ ($j = 2,\dots, m^d$) and given a designated site $c_j\in C_j$, there exists a (time-dependent) Hamiltonian $H_j$ that generates a unitary $U_j$ such that
\begin{align}
	(a\ket{0}+b\ket{1})_{c_j} \ket{\bar 0}_{C_j \setminus c_j}
	\xrightarrow{U_j} a \ket{\bar 0}_{C_j} + b\ket{\bar 1}_{C_j}
\end{align}
for all complex coefficients $a$ and $b$, in time $t_1$ satisfying \cref{eq:t1bound}.
By linearity, this property applies even if $C_j$ is entangled with other hypercubes.
Consequently, backward time evolution under $H_j$ generates $U_j^\dag$, which ``undoes'' the GHZ-like state of the $j$th hypercube:
\begin{align}
	  a \ket{\bar 0}_{C_j} + b \ket{\bar 1}_{C_j}
	\xrightarrow{U_j^\dagger} (a \ket{0}+b\ket{1})_{c_j} \ket{\bar 0}_{C_j \setminus c_j}
\end{align}
for any $a,b$.
In this step, we simultaneously apply $U_j^\dagger$ to $C_j$ for all $j = 2,\dots,m^d$.
These unitaries rotate the state of $C$ to
\begin{align}
	 a\ket{\bar 0}_{C_1} \bigotimes_{j=2}^{m^d} \ket{+}_{c_j}\ket{\bar 0}_{C_j \setminus c_j}
	 + b\ket{\bar 1}_{C_1} \bigotimes_{j=2}^{m^d} \ket{-}_{c_j} \ket{\bar 0}_{C_j \setminus c_j},
\end{align}
where $\ket{\pm} = (\ket{0}\pm\ket{1})/{\sqrt{2}}$.

\textbf{Step 4:} We then apply a Hadamard gate, i.e.
\begin{align}
	\frac{1}{\sqrt{2}}
 	\begin{pmatrix}
 	 		1 & 1\\
 	 		1 &-1
 	 	\end{pmatrix},
 \end{align} to the site $c_j$ of each hypercubes $C_j$, $j = 2,\dots,m^d$.
These Hadamard gates can be implemented arbitrarily fast since we do not assume any constraints on the single-site terms of the Hamiltonian. The state of $C$ by the end of this step is
\begin{align}
	  a\ket{\bar 0}_{C_1} \bigotimes_{j=2}^{m^d} \ket{0}_{c_j} \ket{\bar 0}_{C_j \setminus c_j}
	 + b\ket{\bar 1}_{C_1} \bigotimes_{j=2}^{m^d} \ket{1}_{c_j} \ket{\bar 0}_{C_j \setminus c_j}.
\end{align}

\textbf{Step 5:} Finally, we apply $U_j$ again to each hypercube $C_j$ ($j=2,\dots, m^d$) to obtain the desired GHZ-like state:
\begin{align}
	 a\ket{\bar 0}_{C_1} \bigotimes_{j=2}^{m^d}  \ket{\bar 0}_{C_j }
	  +b\ket{\bar 1}_{C_1} \bigotimes_{j=2}^{m^d} \ket{\bar 1}_{C_j} = \GHZab_C.
\end{align}

At the end of this routine, we have implemented the unitary satisfying \cref{eq:UGHZ} in time
\begin{align}
	t = 3t_1 + t_2
	= 3t_1 + \pi d^{\alpha/2} m^\alpha r_1^{\alpha-2d}. \label{eq:Tsum}
\end{align}
Solving this recurrence relation for three different ranges of $\alpha$ yields the scalings given in \cref{eq:Tbound}.

\section{Analysis of crosstalk interactions during the Tran protocol}
\label{app:crosstalk_analysis}
In this section, we estimate the number of colors required to upper bound the potential error incurred by the crosstalk interactions from creating GHZ states in parallel during the Tran protocol. The analysis assumes that one of the procedures outlined in \cref{sec:crosstalk} of the main text has already been used to cancel out interactions between plaquettes with different color labels.

To estimate the total amount of crosstalk error in the worst case, we first compute the total possible crosstalk interactions between all plaquettes of the same color and multiply them by the timescales required to create the GHZ states.
We then sum over the sizes of the plaquettes required by different steps in the protocol and multiply by the number of colors, $n$, in order to obtain an estimate of the worst-case error incurred over the course of the whole protocol.

For a given step of the protocol with a plaquette size of $L$, we define the worst-case crosstalk error $\eps(L)$ to be
\begin{align}
   \eps(L) = n t_\text{GHZ}(L) \|H_\text{crosstalk}(L,R)\|,
\end{align}
where $t_\text{GHZ}(L)$ is the time required to create a GHZ state of size $L$ by merging GHZ states from the previous step and $\|H_\text{crosstalk}(L,R)\| = \max_{C_i} \sum_{C_j} \|H_{C_i C_j}\|$, where $C_i$ and $C_j$ are plaquettes of the same color separated by a distance of at least $R$, is the largest possible crosstalk interaction between plaquettes of the same color for fixed $L$ and $R$.

We observe that $\|H_{C_i C_j}\| \le \frac{|C_i||C_j|}{d(C_i,C_j)^\al} = \frac{L^{2d}}{d(C_i,C_j)^\al}$, where $d(C_i,C_j)$ is the distance between the sets $C_i$ and $C_j$.
Bounding the sum in $\|H_\text{crosstalk}(L,R)\|$ by an integral, we have:
\begin{align}
	\eps(L) \lesssim n t_\text{GHZ}(L) \int_1^{r/R} d^d\vec x \frac{L^{2d}}{(R|\vec x|)^\al},
\end{align}
where $r$ is the length scale corresponding to the final (largest) GHZ state being created.
For all $\al > d$, the integral can be upper bounded by a constant times $L^{2d}/R^\al$.

The time $t_\text{GHZ}$ for each GHZ-creation step has the same scaling as in \cref{eq:Tbound}, which we reproduce here for ease of reference:
\begin{align}
	t_\text{GHZ}(L)  \leq K_\alpha\times\begin{cases}
		\log^{\kappa_\alpha} L & \text{if } d< \alpha < 2d,\\
		e^{\gamma \sqrt{\log L}} &\text{if } \alpha = 2d, \\
		L^{\alpha-2d} & \text{if }2d< \alpha\leq 2d+1.
	\end{cases}
\end{align}

For $\al < 2d$, $m \propto L^{\lambda-1}$, where $\lambda = 2d/\al$.
To compute the total crosstalk error, it is necessary to sum over all length scales $L_i$ from $r_0$ to $r$.
It can be observed that $L_i$ grows doubly exponentially in the number of time steps $i$, due to the relation $L_{i+1}=mL_i = L_i^{\lambda}$, where we note that $\lambda>1$.
Since the largest crosstalk error is incurred when $L_i=r$, the total crosstalk error $\eps_\text{tot}$ can be upper bounded by the maximum number of recursive steps $i_\text{max}$ times the largest crosstalk error $\eps(r)$ as follows:
\begin{equation}
\begin{aligned}
	\eps_\text{tot} &= \sum_{i} \eps(L_i) = \sum_{i = 0}^{i_\text{max}} \eps(r_0^{\lambda^i}) <  i_\text{max} \eps(r) \\ &\lesssim \log\log(r) \log^{\kappa_\alpha}(r) \frac{r^{2d-\al}}{n^{\al/d-1}},
\end{aligned}
\end{equation}
where $i_\text{max} = \log\log(r)/(\log({\lambda}){\log\log(r_0)})$.
So maintaining constant crosstalk error requires $n \gtrsim \polylog(r) r^{d(2d-\alpha)/(\alpha-d)}$ colors.
This leads to the first line in \cref{eq:colors}.

For $\al = 2d$, the crosstalk error is given by
\begin{align}
	\eps(L) \lesssim e^{\gamma \sqrt{\log L}} \frac{L^{2d-\al}}{n^{\al/d-1}} = \frac{e^{\gamma \sqrt{\log L}}} {n}.
\end{align}
Since the scaling of $m$ is given by $m \propto e^{\frac{\gamma}{2d} \sqrt{\log L}}$, where $\gamma = 3\sqrt{d}$, we have that $m(L) > m'$ for some constant $m' = O(1)$.
Thus, the scaling of $L_i$ throughout the recursive protocol is at least geometric, so the number of steps will be at most $\log(r)$. So the overall crosstalk error can be bounded by
\begin{equation}
\begin{aligned}
	\eps_\text{tot} &= \sum_{i} \eps(L_i) = \sum_{i = 0}^{i_\text{max}} \eps\left((m')^i r_0\right) < i_\text{max} \eps(r) \\
    &\lesssim  \log(r)\frac{e^{{\gamma}\sqrt{\log r}}}{n},
\end{aligned}
\end{equation}
where $i_\text{max} = \log_{m'}(r/r_0)$. So to cancel out the crosstalk error, the scaling of $n$ with $r$ must be at least $n \gtrsim\log(r)e^{\gamma\sqrt{\log r}}$, which is faster than any logarithmic function of $r$, and reproduces the second line of \cref{eq:colors}.

Finally, for $\al \in (2d,2d+1)$,  the scaling of $L_i$ increases geometrically with $i$ and given by $L_{i+1} = cL_i^2$ for some constant $c$.
So the crosstalk error is given by
\begin{align}
   \eps_\text{tot} &= \sum_{i} \eps(L_i) = \sum_{i=0}^{i_\text{max}} \eps\left(c^ir_0 \right)\\
 &=\sum_{i=0}^{\log_m \frac{r}{r_0}} (c^i r_0 )^{\al-2d}  \frac{( c^i r_0)^{2d-\alpha}}{n^{\alpha/d-1}} \lesssim \frac{\log(r)}{n^{\al/d-1}},
\end{align}
so the scaling of the colors must be $n\gtrsim \log(r)^{d/(\al-d)}$, confirming the last line in \cref{eq:colors}.

\section{Polar molecule interactions}
\label{app:polar}
In this section, we present the details of the derivation for the polar molecule Ising interactions using a single dressing field at a time. At the end of this section, we discuss the dressing necessary to realize asymmetric blockade with polar molecules.

\begin{figure*}
    \centering
    \includegraphics[scale=.7]{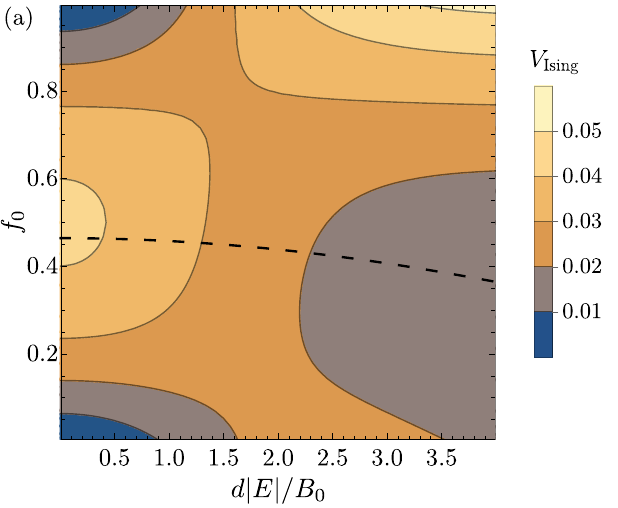}
    \qquad \qquad
    \includegraphics[scale=.7]{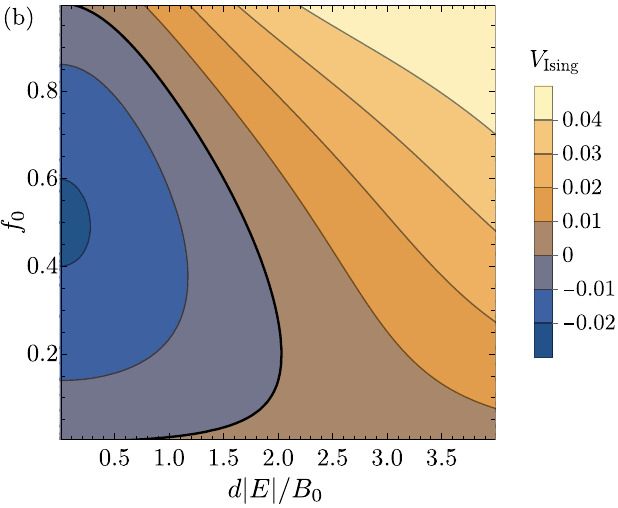}
    \caption{Ising interactions for scheme (a) and (b) as a function of $f_0$ and $d|E|/B_0$. The equal superposition $f_0 =1/2$ is discussed in the main text. In (a), the dashed line denotes when the effective dipole moments of $|0\rangle$ and $|1\rangle$ are equal. In (b), the thick black line indicates when the Ising interactions change sign.}
    \label{fig:Isingab}
\end{figure*}

First, we remark on which interaction terms remain resonant and time-independent in the rotating frame of the microwave field. For schemes (a) and (b), these are

\begin{widetext}
\begin{subequations}
\label{eq:Hm}
\begin{equation}
\begin{aligned}
    V_\text{a} = \frac{1}{r_{ij}^3} \big[& \left(\mu_0 |\phi_{0,0} \rangle \langle  \phi_{0,0}|
    +  \mu_1 |\phi_{1,0}  \rangle \langle \phi_{1,0}|
     +  \mu_2 |\phi_{2,-2}  \rangle \langle \phi_{2,-2}|\right) \times \left(\mu_0 |\phi_{0,0} \rangle \langle  \phi_{0,0}|
    +  \mu_1 |\phi_{1,0}  \rangle \langle \phi_{1,0}|
     +  \mu_2 |\phi_{2,-2}  \rangle \langle \phi_{2,-2}|\right) +~  \\
    & \mu_{01}^2 |\phi_{0,0} \phi_{1,0} \rangle \langle \phi_{1,0} \phi_{0,0}| + \mu_{01}^2 |\phi_{1,0} \phi_{0,0} \rangle \langle \phi_{0,0} \phi_{1,0}| \big],
\end{aligned}
\end{equation}
\begin{equation}
\begin{aligned}
    V_\text{b} = \frac{1}{r_{ij}^3} \big[& \left(\mu_0 |\phi_{0,0} \rangle \langle  \phi_{0,0}|
    +  \mu_1' |\phi_{1,1}  \rangle \langle \phi_{1,1}|
     +  \mu_2 |\phi_{2,-2}  \rangle \langle \phi_{2,-2}|\right) \times \left(\mu_0 |\phi_{0,0} \rangle \langle  \phi_{0,0}|
    +  \mu_1' |\phi_{1,1}  \rangle \langle \phi_{1,1}|
     +  \mu_2 |\phi_{2,-2}  \rangle \langle \phi_{2,-2}|\right) +~  \\
    & (-\mu_{01}'^2 |\phi_{0,0} \phi_{1,1} \rangle \langle \phi_{1,1} \phi_{0,0}| - \mu_{01}'^2 |\phi_{1,1} \phi_{0,0} \rangle \langle \phi_{0,0} \phi_{1,1}|
    )/2 \big],
\end{aligned}
\end{equation}
\end{subequations}
\end{widetext}
respectively, where the flip-flop interactions in the last line of equation equation differ in sign and magnitude.

We are interested in dressed states of the form $|0/0'\rangle \equiv \sqrt{f_0} |\phi_{0,0}\rangle + \sqrt{1-f_0} |\phi_{1,0/1} \rangle$ depending on the scheme, where $f_0$ captures the dressed state population of the $|\phi_{0,0}\rangle$ state and the relative phase of the superposition does not affect the interactions. We evaluate the interactions in the basis of interest for scheme (a)
\begin{subequations}
    \begin{equation}
        \langle 0 0| V_a | 0 0 \rangle \propto [f_0 \mu_0 + (1-f_0) \mu_1]^2 + 2 f_0 (1-f_0) \mu_{01}^2,
    \end{equation}
    \begin{equation}
        \langle 0 1 | V_a | 0 1 \rangle \propto [f_0 \mu_0 + (1-f_0) \mu_1] \mu_2 ,
    \end{equation}
    \begin{equation}
        \langle 1 1| V_a | 1 1 \rangle \propto \mu_2^2,
    \end{equation}
\end{subequations}
and for scheme (b)
\begin{subequations}
    \begin{equation}
        \langle 0' 0'| V_b | 0' 0' \rangle \propto [f_0 \mu_0 + (1-f_0) \mu_1']^2 - f_0 (1-f_0) \mu_{01}'^2,
    \end{equation}
    \begin{equation}
        \langle 0' 1 | V_b | 0' 1 \rangle \propto [f_0 \mu_0 + (1-f_0) \mu_1'] \mu_2,
    \end{equation}
    \begin{equation}
        \langle 1 1| V_b | 1 1 \rangle \propto \mu_2^2,
    \end{equation}
\end{subequations}
where the proportionality corresponds to the $r_{ij}$-dependence we have dropped for simplicity. These terms correspond to the $V_0, V_1, V_{01}$ terms (and their primed counterparts) in the main text up to a factor of 4.

In Fig.~\ref{fig:Isingab}, we plot the strength of the Ising interactions as a function of $f_0$ and $d|E|/B_0$. We illustrate the parameters where the effective dipole moments of $|0\rangle, |1\rangle$ are equal for scheme (a) and the region where negative Ising interactions are realized for scheme (b). In the absence of an electric field, the interactions are strongest for both schemes when $f_0 = 1/2$, corresponding to an equal superposition. For scheme (a), the interactions can in principle be increased at sufficiently large electric fields $d|E|/B_0 \gtrsim 2$. For scheme (b), the strongest negative interactions occur at $E = 0$, and increasing the electric field only shifts the interactions towards positive values.

\subsection*{Asymmetric blockade}

In this section, we present some of the details for the dressing parameters needed to realize asymmetric blockade. This is largely reproduced from Ref.~\cite{Young2021}, so the discussion of the dressing parameters applies to Rydberg atoms as well, so long as the drives are not sufficiently strong to cause coupling to additional states.

First, we define $\mu_\pi$, $\mu_\sigma$ to be the transition dipole moments associated with the driven $\pi$ and $\sigma$ transitions, respectively, with the ratio $\mathcal{M} \equiv \mu_\pi/\mu_\sigma$. Additionally, we have the relation $c_\pi = \mathcal{C}/\sqrt{|2 \mathcal{M}^2-1|}$. For a given choice of $\Omega_\pi$, the other dressing parameters are
\begin{subequations}
\begin{equation}
    \Omega_\sigma = \sqrt{2} \mathcal{M} \frac{1 + c_\pi^2 (1- 2 \mathcal{M}^2)}{1-c_\pi^2(1-2 \mathcal{M}^2)} \Omega_\pi,
\end{equation}
\begin{equation}
    \Delta_\pi = \frac{1 - 2 c_\pi^2 + \mathcal{C}^4}{c_\pi (1-c_\pi^2(1-2  \mathcal{M}^2))} \Omega_\pi,
\end{equation}
\begin{equation}
    \Delta_\sigma = \frac{1- 4 c_\pi^2 \mathcal{M}^2 + \mathcal{C}^4}{c_\pi (1-c_\pi^2(1-2  \mathcal{M}^2))}\Omega_\pi,
\end{equation}
\end{subequations}
which are naturally all proportional to $\Omega_\pi$. At $\mathcal{C} = 1$, the dipolar c-t interactions are maximized with a value of
\begin{equation}
    C_3^\text{max} = \min \left(\frac{\mu_\pi^2}{\mu_\sigma^2/2}, \frac{\mu_\sigma^2/2}{\mu_\pi^2}\right) (\mu_\pi^2-\mu_\sigma^2/2),
\end{equation}
although, as discussed in the main text, the two dressed states $|c\rangle, |t\rangle$ are degenerate when this is realized.

Next, we discuss the vdW interactions in the context of polar molecules. As indicated by the discussion in the main text, the vdW interactions for polar molecules are less complicated than for Rydberg atoms. This is because the transition energy differences are typically on the order of GHz, much larger than the dressing and interaction scales. As a result, only the resonant dipolar interactions [e.g.,~the flip-flop terms in Eq.~(\ref{eq:Hm})] are relevant to the vdW interactions, so a Floquet approach is not needed to deal with the multiple rotating frames as it is for Rydberg atoms.

In our example, these interactions are the flip-flop interactions between the $N=0$ and $N=1$ states as well as the flip-flop interactions between the $N=1$ and $N=2$ states, which we capture via a dipolar interaction Hamiltonian $H_\text{dd}$. Furthermore, the drives also couple additional transitions, such as $|1,1\rangle \to |2,2\rangle$, among others. In light of this, we utilize the same rotating frame for all $N=2$ states as the $|2,1\rangle$ state, including the uncoupled states $|2,-2\rangle,|2,-1\rangle$, which allows us to write $H_\text{dd}$ in a fully time-independent form. As a result, in the dressed basis, we have the usual expression for the vdW interactions:
\begin{equation}
    V_\text{vdW}^{(c)} = \sum_{k \neq cc} \frac{|\langle k| H_\text{dd}| cc \rangle|^2}{ 2 E_c - E_k},
\end{equation}
where $k$ are all other dressed state pairs in the basis and $E_k$ the corresponding dressed state energies in the rotating frame.

\bibliography{state_transfer_expt_bbt2}

\begin{thebibliography}{48}%
\makeatletter
\providecommand \@ifxundefined [1]{%
 \@ifx{#1\undefined}
}%
\providecommand \@ifnum [1]{%
 \ifnum #1\expandafter \@firstoftwo
 \else \expandafter \@secondoftwo
 \fi
}%
\providecommand \@ifx [1]{%
 \ifx #1\expandafter \@firstoftwo
 \else \expandafter \@secondoftwo
 \fi
}%
\providecommand \natexlab [1]{#1}%
\providecommand \enquote  [1]{``#1''}%
\providecommand \bibnamefont  [1]{#1}%
\providecommand \bibfnamefont [1]{#1}%
\providecommand \citenamefont [1]{#1}%
\providecommand \href@noop [0]{\@secondoftwo}%
\providecommand \href [0]{\begingroup \@sanitize@url \@href}%
\providecommand \@href[1]{\@@startlink{#1}\@@href}%
\providecommand \@@href[1]{\endgroup#1\@@endlink}%
\providecommand \@sanitize@url [0]{\catcode `\\12\catcode `\$12\catcode
  `\&12\catcode `\#12\catcode `\^12\catcode `\_12\catcode `\%12\relax}%
\providecommand \@@startlink[1]{}%
\providecommand \@@endlink[0]{}%
\providecommand \url  [0]{\begingroup\@sanitize@url \@url }%
\providecommand \@url [1]{\endgroup\@href {#1}{\urlprefix }}%
\providecommand \urlprefix  [0]{URL }%
\providecommand \Eprint [0]{\href }%
\providecommand \doibase [0]{https://doi.org/}%
\providecommand \selectlanguage [0]{\@gobble}%
\providecommand \bibinfo  [0]{\@secondoftwo}%
\providecommand \bibfield  [0]{\@secondoftwo}%
\providecommand \translation [1]{[#1]}%
\providecommand \BibitemOpen [0]{}%
\providecommand \bibitemStop [0]{}%
\providecommand \bibitemNoStop [0]{.\EOS\space}%
\providecommand \EOS [0]{\spacefactor3000\relax}%
\providecommand \BibitemShut  [1]{\csname bibitem#1\endcsname}%
\let\auto@bib@innerbib\@empty
\bibitem [{\citenamefont {Endres}\ \emph {et~al.}(2016)\citenamefont {Endres},
  \citenamefont {Bernien}, \citenamefont {Keesling}, \citenamefont {Levine},
  \citenamefont {Anschuetz}, \citenamefont {Krajenbrink}, \citenamefont
  {Senko}, \citenamefont {Vuletic}, \citenamefont {Greiner},\ and\
  \citenamefont {Lukin}}]{Endres2016}%
  \BibitemOpen
  \bibfield  {author} {\bibinfo {author} {\bibfnamefont {M.}~\bibnamefont
  {Endres}}, \bibinfo {author} {\bibfnamefont {H.}~\bibnamefont {Bernien}},
  \bibinfo {author} {\bibfnamefont {A.}~\bibnamefont {Keesling}}, \bibinfo
  {author} {\bibfnamefont {H.}~\bibnamefont {Levine}}, \bibinfo {author}
  {\bibfnamefont {E.~R.}\ \bibnamefont {Anschuetz}}, \bibinfo {author}
  {\bibfnamefont {A.}~\bibnamefont {Krajenbrink}}, \bibinfo {author}
  {\bibfnamefont {C.}~\bibnamefont {Senko}}, \bibinfo {author} {\bibfnamefont
  {V.}~\bibnamefont {Vuletic}}, \bibinfo {author} {\bibfnamefont
  {M.}~\bibnamefont {Greiner}},\ and\ \bibinfo {author} {\bibfnamefont {M.~D.}\
  \bibnamefont {Lukin}},\ }\bibfield  {title} {\bibinfo {title} {Atom-by-atom
  assembly of defect-free one-dimensional cold atom arrays},\ }\href
  {https://doi.org/10.1126/science.aah3752} {\bibfield  {journal} {\bibinfo
  {journal} {Science}\ }\textbf {\bibinfo {volume} {354}},\ \bibinfo {pages}
  {1024} (\bibinfo {year} {2016})}\BibitemShut {NoStop}%
\bibitem [{\citenamefont {Barredo}\ \emph {et~al.}(2018)\citenamefont
  {Barredo}, \citenamefont {Lienhard}, \citenamefont {{de L{\'e}s{\'e}leuc}},
  \citenamefont {Lahaye},\ and\ \citenamefont {Browaeys}}]{Barredo2018}%
  \BibitemOpen
  \bibfield  {author} {\bibinfo {author} {\bibfnamefont {D.}~\bibnamefont
  {Barredo}}, \bibinfo {author} {\bibfnamefont {V.}~\bibnamefont {Lienhard}},
  \bibinfo {author} {\bibfnamefont {S.}~\bibnamefont {{de L{\'e}s{\'e}leuc}}},
  \bibinfo {author} {\bibfnamefont {T.}~\bibnamefont {Lahaye}},\ and\ \bibinfo
  {author} {\bibfnamefont {A.}~\bibnamefont {Browaeys}},\ }\bibfield  {title}
  {\bibinfo {title} {Synthetic three-dimensional atomic structures assembled
  atom by atom},\ }\href {https://doi.org/10.1038/s41586-018-0450-2} {\bibfield
   {journal} {\bibinfo  {journal} {Nature}\ }\textbf {\bibinfo {volume}
  {561}},\ \bibinfo {pages} {79} (\bibinfo {year} {2018})}\BibitemShut
  {NoStop}%
\bibitem [{\citenamefont {Browaeys}\ and\ \citenamefont
  {Lahaye}(2020)}]{Browaeys2020}%
  \BibitemOpen
  \bibfield  {author} {\bibinfo {author} {\bibfnamefont {A.}~\bibnamefont
  {Browaeys}}\ and\ \bibinfo {author} {\bibfnamefont {T.}~\bibnamefont
  {Lahaye}},\ }\bibfield  {title} {\bibinfo {title} {Many-body physics with
  individually controlled {{Rydberg}} atoms},\ }\href
  {https://doi.org/10.1038/s41567-019-0733-z} {\bibfield  {journal} {\bibinfo
  {journal} {Nat. Phys.}\ }\textbf {\bibinfo {volume} {16}},\ \bibinfo {pages}
  {132} (\bibinfo {year} {2020})}\BibitemShut {NoStop}%
\bibitem [{\citenamefont {Kaufman}\ and\ \citenamefont
  {Ni}(2021)}]{Kaufman2021}%
  \BibitemOpen
  \bibfield  {author} {\bibinfo {author} {\bibfnamefont {A.~M.}\ \bibnamefont
  {Kaufman}}\ and\ \bibinfo {author} {\bibfnamefont {K.-K.}\ \bibnamefont
  {Ni}},\ }\bibfield  {title} {\bibinfo {title} {Quantum science with optical
  tweezer arrays of ultracold atoms and molecules},\ }\href
  {https://doi.org/10.1038/s41567-021-01357-2} {\bibfield  {journal} {\bibinfo
  {journal} {Nat. Phys.}\ }\textbf {\bibinfo {volume} {17}},\ \bibinfo {pages}
  {1324} (\bibinfo {year} {2021})}\BibitemShut {NoStop}%
\bibitem [{\citenamefont {Bloch}\ \emph {et~al.}(2023)\citenamefont {Bloch},
  \citenamefont {Hofer}, \citenamefont {Cohen}, \citenamefont {Browaeys},\ and\
  \citenamefont {{Ferrier-Barbut}}}]{Bloch2023}%
  \BibitemOpen
  \bibfield  {author} {\bibinfo {author} {\bibfnamefont {D.}~\bibnamefont
  {Bloch}}, \bibinfo {author} {\bibfnamefont {B.}~\bibnamefont {Hofer}},
  \bibinfo {author} {\bibfnamefont {S.~R.}\ \bibnamefont {Cohen}}, \bibinfo
  {author} {\bibfnamefont {A.}~\bibnamefont {Browaeys}},\ and\ \bibinfo
  {author} {\bibfnamefont {I.}~\bibnamefont {{Ferrier-Barbut}}},\ }\bibfield
  {title} {\bibinfo {title} {Trapping and {{Imaging Single Dysprosium Atoms}}
  in {{Optical Tweezer Arrays}}},\ }\href
  {https://doi.org/10.1103/PhysRevLett.131.203401} {\bibfield  {journal}
  {\bibinfo  {journal} {Phys. Rev. Lett.}\ }\textbf {\bibinfo {volume} {131}},\
  \bibinfo {pages} {203401} (\bibinfo {year} {2023})}\BibitemShut {NoStop}%
\bibitem [{\citenamefont {Fraxanet}\ \emph {et~al.}(2022)\citenamefont
  {Fraxanet}, \citenamefont {{Gonz{\'a}lez-Cuadra}}, \citenamefont {Pfau},
  \citenamefont {Lewenstein}, \citenamefont {Langen},\ and\ \citenamefont
  {Barbiero}}]{Fraxanet2022}%
  \BibitemOpen
  \bibfield  {author} {\bibinfo {author} {\bibfnamefont {J.}~\bibnamefont
  {Fraxanet}}, \bibinfo {author} {\bibfnamefont {D.}~\bibnamefont
  {{Gonz{\'a}lez-Cuadra}}}, \bibinfo {author} {\bibfnamefont {T.}~\bibnamefont
  {Pfau}}, \bibinfo {author} {\bibfnamefont {M.}~\bibnamefont {Lewenstein}},
  \bibinfo {author} {\bibfnamefont {T.}~\bibnamefont {Langen}},\ and\ \bibinfo
  {author} {\bibfnamefont {L.}~\bibnamefont {Barbiero}},\ }\bibfield  {title}
  {\bibinfo {title} {Topological {{Quantum Critical Points}} in the {{Extended
  Bose-Hubbard Model}}},\ }\href
  {https://doi.org/10.1103/PhysRevLett.128.043402} {\bibfield  {journal}
  {\bibinfo  {journal} {Phys. Rev. Lett.}\ }\textbf {\bibinfo {volume} {128}},\
  \bibinfo {pages} {043402} (\bibinfo {year} {2022})}\BibitemShut {NoStop}%
\bibitem [{\citenamefont {Gross}\ and\ \citenamefont {Bakr}(2021)}]{Gross2021}%
  \BibitemOpen
  \bibfield  {author} {\bibinfo {author} {\bibfnamefont {C.}~\bibnamefont
  {Gross}}\ and\ \bibinfo {author} {\bibfnamefont {W.~S.}\ \bibnamefont
  {Bakr}},\ }\bibfield  {title} {\bibinfo {title} {Quantum gas microscopy for
  single atom and spin detection},\ }\href
  {https://doi.org/10.1038/s41567-021-01370-5} {\bibfield  {journal} {\bibinfo
  {journal} {Nat. Phys.}\ }\textbf {\bibinfo {volume} {17}},\ \bibinfo {pages}
  {1316} (\bibinfo {year} {2021})}\BibitemShut {NoStop}%
\bibitem [{\citenamefont {Anich}\ \emph {et~al.}(2023)\citenamefont {Anich},
  \citenamefont {Grimm},\ and\ \citenamefont {Kirilov}}]{Anich2023}%
  \BibitemOpen
  \bibfield  {author} {\bibinfo {author} {\bibfnamefont {G.}~\bibnamefont
  {Anich}}, \bibinfo {author} {\bibfnamefont {R.}~\bibnamefont {Grimm}},\ and\
  \bibinfo {author} {\bibfnamefont {E.}~\bibnamefont {Kirilov}},\ }\href@noop
  {} {\bibinfo {title} {Comprehensive {{Characterization}} of a
  {{State-of-the-Art Apparatus}} for {{Cold Electromagnetic Dysprosium
  Dipoles}}}} (\bibinfo {year} {2023}),\ \Eprint
  {https://arxiv.org/abs/2304.12844} {arxiv:2304.12844} \BibitemShut {NoStop}%
\bibitem [{\citenamefont {Su}\ \emph {et~al.}(2023)\citenamefont {Su},
  \citenamefont {Douglas}, \citenamefont {Szurek}, \citenamefont {Groth},
  \citenamefont {Ozturk}, \citenamefont {Krahn}, \citenamefont {H{\'e}bert},
  \citenamefont {Phelps}, \citenamefont {Ebadi}, \citenamefont {Dickerson},
  \citenamefont {Ferlaino}, \citenamefont {Markovi{\'c}},\ and\ \citenamefont
  {Greiner}}]{Su2023}%
  \BibitemOpen
  \bibfield  {author} {\bibinfo {author} {\bibfnamefont {L.}~\bibnamefont
  {Su}}, \bibinfo {author} {\bibfnamefont {A.}~\bibnamefont {Douglas}},
  \bibinfo {author} {\bibfnamefont {M.}~\bibnamefont {Szurek}}, \bibinfo
  {author} {\bibfnamefont {R.}~\bibnamefont {Groth}}, \bibinfo {author}
  {\bibfnamefont {S.~F.}\ \bibnamefont {Ozturk}}, \bibinfo {author}
  {\bibfnamefont {A.}~\bibnamefont {Krahn}}, \bibinfo {author} {\bibfnamefont
  {A.~H.}\ \bibnamefont {H{\'e}bert}}, \bibinfo {author} {\bibfnamefont
  {G.~A.}\ \bibnamefont {Phelps}}, \bibinfo {author} {\bibfnamefont
  {S.}~\bibnamefont {Ebadi}}, \bibinfo {author} {\bibfnamefont
  {S.}~\bibnamefont {Dickerson}}, \bibinfo {author} {\bibfnamefont
  {F.}~\bibnamefont {Ferlaino}}, \bibinfo {author} {\bibfnamefont
  {O.}~\bibnamefont {Markovi{\'c}}},\ and\ \bibinfo {author} {\bibfnamefont
  {M.}~\bibnamefont {Greiner}},\ }\bibfield  {title} {\bibinfo {title} {Dipolar
  quantum solids emerging in a {{Hubbard}} quantum simulator},\ }\href
  {https://doi.org/10.1038/s41586-023-06614-3} {\bibfield  {journal} {\bibinfo
  {journal} {Nature}\ }\textbf {\bibinfo {volume} {622}},\ \bibinfo {pages}
  {724} (\bibinfo {year} {2023})}\BibitemShut {NoStop}%
\bibitem [{\citenamefont {Sohmen}\ \emph {et~al.}(2023)\citenamefont {Sohmen},
  \citenamefont {Mark}, \citenamefont {Greiner},\ and\ \citenamefont
  {Ferlaino}}]{Sohmen2023}%
  \BibitemOpen
  \bibfield  {author} {\bibinfo {author} {\bibfnamefont {M.}~\bibnamefont
  {Sohmen}}, \bibinfo {author} {\bibfnamefont {M.~J.}\ \bibnamefont {Mark}},
  \bibinfo {author} {\bibfnamefont {M.}~\bibnamefont {Greiner}},\ and\ \bibinfo
  {author} {\bibfnamefont {F.}~\bibnamefont {Ferlaino}},\ }\bibfield  {title}
  {\bibinfo {title} {A ship-in-a-bottle quantum gas microscope setup for
  magnetic mixtures},\ }\href {https://doi.org/10.21468/SciPostPhys.15.5.182}
  {\bibfield  {journal} {\bibinfo  {journal} {Scipost Phys}\ }\textbf {\bibinfo
  {volume} {15}},\ \bibinfo {pages} {182} (\bibinfo {year} {2023})}\BibitemShut
  {NoStop}%
\bibitem [{\citenamefont {Chomaz}\ \emph {et~al.}(2022)\citenamefont {Chomaz},
  \citenamefont {{Ferrier-Barbut}}, \citenamefont {Ferlaino}, \citenamefont
  {{Laburthe-Tolra}}, \citenamefont {Lev},\ and\ \citenamefont
  {Pfau}}]{Chomaz2022}%
  \BibitemOpen
  \bibfield  {author} {\bibinfo {author} {\bibfnamefont {L.}~\bibnamefont
  {Chomaz}}, \bibinfo {author} {\bibfnamefont {I.}~\bibnamefont
  {{Ferrier-Barbut}}}, \bibinfo {author} {\bibfnamefont {F.}~\bibnamefont
  {Ferlaino}}, \bibinfo {author} {\bibfnamefont {B.}~\bibnamefont
  {{Laburthe-Tolra}}}, \bibinfo {author} {\bibfnamefont {B.~L.}\ \bibnamefont
  {Lev}},\ and\ \bibinfo {author} {\bibfnamefont {T.}~\bibnamefont {Pfau}},\
  }\bibfield  {title} {\bibinfo {title} {Dipolar physics: A review of
  experiments with magnetic quantum gases},\ }\href
  {https://doi.org/10.1088/1361-6633/aca814} {\bibfield  {journal} {\bibinfo
  {journal} {Rep. Prog. Phys.}\ }\textbf {\bibinfo {volume} {86}},\ \bibinfo
  {pages} {026401} (\bibinfo {year} {2022})}\BibitemShut {NoStop}%
\bibitem [{\citenamefont {Liu}\ \emph {et~al.}(2018)\citenamefont {Liu},
  \citenamefont {Hood}, \citenamefont {Yu}, \citenamefont {Zhang},
  \citenamefont {Hutzler}, \citenamefont {Rosenband},\ and\ \citenamefont
  {Ni}}]{Liu2018}%
  \BibitemOpen
  \bibfield  {author} {\bibinfo {author} {\bibfnamefont {L.~R.}\ \bibnamefont
  {Liu}}, \bibinfo {author} {\bibfnamefont {J.~D.}\ \bibnamefont {Hood}},
  \bibinfo {author} {\bibfnamefont {Y.}~\bibnamefont {Yu}}, \bibinfo {author}
  {\bibfnamefont {J.~T.}\ \bibnamefont {Zhang}}, \bibinfo {author}
  {\bibfnamefont {N.~R.}\ \bibnamefont {Hutzler}}, \bibinfo {author}
  {\bibfnamefont {T.}~\bibnamefont {Rosenband}},\ and\ \bibinfo {author}
  {\bibfnamefont {K.-K.}\ \bibnamefont {Ni}},\ }\bibfield  {title} {\bibinfo
  {title} {Building one molecule from a reservoir of two atoms},\ }\href
  {https://doi.org/10.1126/science.aar7797} {\bibfield  {journal} {\bibinfo
  {journal} {Science}\ }\textbf {\bibinfo {volume} {360}},\ \bibinfo {pages}
  {900} (\bibinfo {year} {2018})}\BibitemShut {NoStop}%
\bibitem [{\citenamefont {Anderegg}\ \emph {et~al.}(2019)\citenamefont
  {Anderegg}, \citenamefont {Cheuk}, \citenamefont {Bao}, \citenamefont
  {Burchesky}, \citenamefont {Ketterle}, \citenamefont {Ni},\ and\
  \citenamefont {{John M. Doyle}}}]{Anderegg2019}%
  \BibitemOpen
  \bibfield  {author} {\bibinfo {author} {\bibfnamefont {L.}~\bibnamefont
  {Anderegg}}, \bibinfo {author} {\bibfnamefont {L.~W.}\ \bibnamefont {Cheuk}},
  \bibinfo {author} {\bibfnamefont {Y.}~\bibnamefont {Bao}}, \bibinfo {author}
  {\bibfnamefont {S.}~\bibnamefont {Burchesky}}, \bibinfo {author}
  {\bibfnamefont {W.}~\bibnamefont {Ketterle}}, \bibinfo {author}
  {\bibfnamefont {K.-K.}\ \bibnamefont {Ni}},\ and\ \bibinfo {author}
  {\bibnamefont {{John M. Doyle}}},\ }\bibfield  {title} {\bibinfo {title} {An
  optical tweezer array of ultracold molecules},\ }\href
  {https://doi.org/10.1126/science.aax1265} {\bibfield  {journal} {\bibinfo
  {journal} {Science}\ }\textbf {\bibinfo {volume} {365}},\ \bibinfo {pages}
  {1156} (\bibinfo {year} {2019})}\BibitemShut {NoStop}%
\bibitem [{\citenamefont {Zhang}\ \emph {et~al.}(2022)\citenamefont {Zhang},
  \citenamefont {Picard}, \citenamefont {Cairncross}, \citenamefont {Wang},
  \citenamefont {Yu}, \citenamefont {Fang},\ and\ \citenamefont
  {Ni}}]{Zhang2022}%
  \BibitemOpen
  \bibfield  {author} {\bibinfo {author} {\bibfnamefont {J.~T.}\ \bibnamefont
  {Zhang}}, \bibinfo {author} {\bibfnamefont {L.~R.~B.}\ \bibnamefont
  {Picard}}, \bibinfo {author} {\bibfnamefont {W.~B.}\ \bibnamefont
  {Cairncross}}, \bibinfo {author} {\bibfnamefont {K.}~\bibnamefont {Wang}},
  \bibinfo {author} {\bibfnamefont {Y.}~\bibnamefont {Yu}}, \bibinfo {author}
  {\bibfnamefont {F.}~\bibnamefont {Fang}},\ and\ \bibinfo {author}
  {\bibfnamefont {K.-K.}\ \bibnamefont {Ni}},\ }\bibfield  {title} {\bibinfo
  {title} {An optical tweezer array of ground-state polar molecules},\ }\href
  {https://doi.org/10.1088/2058-9565/ac676c} {\bibfield  {journal} {\bibinfo
  {journal} {Quantum Sci. Technol.}\ }\textbf {\bibinfo {volume} {7}},\
  \bibinfo {pages} {035006} (\bibinfo {year} {2022})}\BibitemShut {NoStop}%
\bibitem [{\citenamefont {Vilas}\ \emph {et~al.}(2023)\citenamefont {Vilas},
  \citenamefont {Robichaud}, \citenamefont {Hallas}, \citenamefont {Li},
  \citenamefont {Anderegg},\ and\ \citenamefont {Doyle}}]{Vilas2023}%
  \BibitemOpen
  \bibfield  {author} {\bibinfo {author} {\bibfnamefont {N.~B.}\ \bibnamefont
  {Vilas}}, \bibinfo {author} {\bibfnamefont {P.}~\bibnamefont {Robichaud}},
  \bibinfo {author} {\bibfnamefont {C.}~\bibnamefont {Hallas}}, \bibinfo
  {author} {\bibfnamefont {G.~K.}\ \bibnamefont {Li}}, \bibinfo {author}
  {\bibfnamefont {L.}~\bibnamefont {Anderegg}},\ and\ \bibinfo {author}
  {\bibfnamefont {J.~M.}\ \bibnamefont {Doyle}},\ }\href@noop {} {\bibinfo
  {title} {An optical tweezer array of ultracold polyatomic molecules}}
  (\bibinfo {year} {2023}),\ \Eprint {https://arxiv.org/abs/2311.07529}
  {arxiv:2311.07529} \BibitemShut {NoStop}%
\bibitem [{\citenamefont {Covey}\ \emph {et~al.}(2018)\citenamefont {Covey},
  \citenamefont {Marco}, \citenamefont {Acevedo}, \citenamefont {Rey},\ and\
  \citenamefont {Ye}}]{Covey2018}%
  \BibitemOpen
  \bibfield  {author} {\bibinfo {author} {\bibfnamefont {J.~P.}\ \bibnamefont
  {Covey}}, \bibinfo {author} {\bibfnamefont {L.~D.}\ \bibnamefont {Marco}},
  \bibinfo {author} {\bibfnamefont {{\'O}.~L.}\ \bibnamefont {Acevedo}},
  \bibinfo {author} {\bibfnamefont {A.~M.}\ \bibnamefont {Rey}},\ and\ \bibinfo
  {author} {\bibfnamefont {J.}~\bibnamefont {Ye}},\ }\bibfield  {title}
  {\bibinfo {title} {An approach to spin-resolved molecular gas microscopy},\
  }\href {https://doi.org/10.1088/1367-2630/aaba65} {\bibfield  {journal}
  {\bibinfo  {journal} {New J. Phys.}\ }\textbf {\bibinfo {volume} {20}},\
  \bibinfo {pages} {043031} (\bibinfo {year} {2018})}\BibitemShut {NoStop}%
\bibitem [{\citenamefont {Christakis}\ \emph {et~al.}(2023)\citenamefont
  {Christakis}, \citenamefont {Rosenberg}, \citenamefont {Raj}, \citenamefont
  {Chi}, \citenamefont {Morningstar}, \citenamefont {Huse}, \citenamefont
  {Yan},\ and\ \citenamefont {Bakr}}]{Christakis2023}%
  \BibitemOpen
  \bibfield  {author} {\bibinfo {author} {\bibfnamefont {L.}~\bibnamefont
  {Christakis}}, \bibinfo {author} {\bibfnamefont {J.~S.}\ \bibnamefont
  {Rosenberg}}, \bibinfo {author} {\bibfnamefont {R.}~\bibnamefont {Raj}},
  \bibinfo {author} {\bibfnamefont {S.}~\bibnamefont {Chi}}, \bibinfo {author}
  {\bibfnamefont {A.}~\bibnamefont {Morningstar}}, \bibinfo {author}
  {\bibfnamefont {D.~A.}\ \bibnamefont {Huse}}, \bibinfo {author}
  {\bibfnamefont {Z.~Z.}\ \bibnamefont {Yan}},\ and\ \bibinfo {author}
  {\bibfnamefont {W.~S.}\ \bibnamefont {Bakr}},\ }\bibfield  {title} {\bibinfo
  {title} {Probing site-resolved correlations in a spin system of ultracold
  molecules},\ }\href {https://doi.org/10.1038/s41586-022-05558-4} {\bibfield
  {journal} {\bibinfo  {journal} {Nature}\ }\textbf {\bibinfo {volume} {614}},\
  \bibinfo {pages} {64} (\bibinfo {year} {2023})}\BibitemShut {NoStop}%
\bibitem [{\citenamefont {Rosenberg}\ \emph {et~al.}(2022)\citenamefont
  {Rosenberg}, \citenamefont {Christakis}, \citenamefont {{Guardado-Sanchez}},
  \citenamefont {Yan},\ and\ \citenamefont {Bakr}}]{Rosenberg2022}%
  \BibitemOpen
  \bibfield  {author} {\bibinfo {author} {\bibfnamefont {J.~S.}\ \bibnamefont
  {Rosenberg}}, \bibinfo {author} {\bibfnamefont {L.}~\bibnamefont
  {Christakis}}, \bibinfo {author} {\bibfnamefont {E.}~\bibnamefont
  {{Guardado-Sanchez}}}, \bibinfo {author} {\bibfnamefont {Z.~Z.}\ \bibnamefont
  {Yan}},\ and\ \bibinfo {author} {\bibfnamefont {W.~S.}\ \bibnamefont
  {Bakr}},\ }\bibfield  {title} {\bibinfo {title} {Observation of the {{Hanbury
  Brown}}{\textendash}{{Twiss}} effect with ultracold molecules},\ }\href
  {https://doi.org/10.1038/s41567-022-01695-9} {\bibfield  {journal} {\bibinfo
  {journal} {Nat. Phys.}\ }\textbf {\bibinfo {volume} {18}},\ \bibinfo {pages}
  {1062} (\bibinfo {year} {2022})}\BibitemShut {NoStop}%
\bibitem [{\citenamefont {Tobias}\ \emph {et~al.}(2022)\citenamefont {Tobias},
  \citenamefont {Matsuda}, \citenamefont {Li}, \citenamefont {Miller},
  \citenamefont {Carroll}, \citenamefont {Bilitewski}, \citenamefont {Rey},\
  and\ \citenamefont {{Jun Ye}}}]{Tobias2022}%
  \BibitemOpen
  \bibfield  {author} {\bibinfo {author} {\bibfnamefont {W.~G.}\ \bibnamefont
  {Tobias}}, \bibinfo {author} {\bibfnamefont {K.}~\bibnamefont {Matsuda}},
  \bibinfo {author} {\bibfnamefont {J.-R.}\ \bibnamefont {Li}}, \bibinfo
  {author} {\bibfnamefont {C.}~\bibnamefont {Miller}}, \bibinfo {author}
  {\bibfnamefont {A.~N.}\ \bibnamefont {Carroll}}, \bibinfo {author}
  {\bibfnamefont {T.}~\bibnamefont {Bilitewski}}, \bibinfo {author}
  {\bibfnamefont {A.~M.}\ \bibnamefont {Rey}},\ and\ \bibinfo {author}
  {\bibnamefont {{Jun Ye}}},\ }\bibfield  {title} {\bibinfo {title} {Reactions
  between layer-resolved molecules mediated by dipolar spin exchange},\ }\href
  {https://doi.org/10.1126/science.abn8525} {\bibfield  {journal} {\bibinfo
  {journal} {Science}\ }\textbf {\bibinfo {volume} {375}},\ \bibinfo {pages}
  {1299} (\bibinfo {year} {2022})}\BibitemShut {NoStop}%
\bibitem [{\citenamefont {Levine}\ \emph {et~al.}(2018)\citenamefont {Levine},
  \citenamefont {Keesling}, \citenamefont {Omran}, \citenamefont {Bernien},
  \citenamefont {Schwartz}, \citenamefont {Zibrov}, \citenamefont {Endres},
  \citenamefont {Greiner}, \citenamefont {Vuleti{\'c}},\ and\ \citenamefont
  {Lukin}}]{Levine2018}%
  \BibitemOpen
  \bibfield  {author} {\bibinfo {author} {\bibfnamefont {H.}~\bibnamefont
  {Levine}}, \bibinfo {author} {\bibfnamefont {A.}~\bibnamefont {Keesling}},
  \bibinfo {author} {\bibfnamefont {A.}~\bibnamefont {Omran}}, \bibinfo
  {author} {\bibfnamefont {H.}~\bibnamefont {Bernien}}, \bibinfo {author}
  {\bibfnamefont {S.}~\bibnamefont {Schwartz}}, \bibinfo {author}
  {\bibfnamefont {A.~S.}\ \bibnamefont {Zibrov}}, \bibinfo {author}
  {\bibfnamefont {M.}~\bibnamefont {Endres}}, \bibinfo {author} {\bibfnamefont
  {M.}~\bibnamefont {Greiner}}, \bibinfo {author} {\bibfnamefont
  {V.}~\bibnamefont {Vuleti{\'c}}},\ and\ \bibinfo {author} {\bibfnamefont
  {M.~D.}\ \bibnamefont {Lukin}},\ }\bibfield  {title} {\bibinfo {title}
  {High-{{Fidelity Control}} and {{Entanglement}} of {{Rydberg-Atom Qubits}}},\
  }\href {https://doi.org/10.1103/PhysRevLett.121.123603} {\bibfield  {journal}
  {\bibinfo  {journal} {Phys. Rev. Lett.}\ }\textbf {\bibinfo {volume} {121}},\
  \bibinfo {pages} {123603} (\bibinfo {year} {2018})}\BibitemShut {NoStop}%
\bibitem [{\citenamefont {Levine}\ \emph {et~al.}(2019)\citenamefont {Levine},
  \citenamefont {Keesling}, \citenamefont {Semeghini}, \citenamefont {Omran},
  \citenamefont {Wang}, \citenamefont {Ebadi}, \citenamefont {Bernien},
  \citenamefont {Greiner}, \citenamefont {Vuleti{\'c}}, \citenamefont
  {Pichler},\ and\ \citenamefont {Lukin}}]{Levine2019}%
  \BibitemOpen
  \bibfield  {author} {\bibinfo {author} {\bibfnamefont {H.}~\bibnamefont
  {Levine}}, \bibinfo {author} {\bibfnamefont {A.}~\bibnamefont {Keesling}},
  \bibinfo {author} {\bibfnamefont {G.}~\bibnamefont {Semeghini}}, \bibinfo
  {author} {\bibfnamefont {A.}~\bibnamefont {Omran}}, \bibinfo {author}
  {\bibfnamefont {T.~T.}\ \bibnamefont {Wang}}, \bibinfo {author}
  {\bibfnamefont {S.}~\bibnamefont {Ebadi}}, \bibinfo {author} {\bibfnamefont
  {H.}~\bibnamefont {Bernien}}, \bibinfo {author} {\bibfnamefont
  {M.}~\bibnamefont {Greiner}}, \bibinfo {author} {\bibfnamefont
  {V.}~\bibnamefont {Vuleti{\'c}}}, \bibinfo {author} {\bibfnamefont
  {H.}~\bibnamefont {Pichler}},\ and\ \bibinfo {author} {\bibfnamefont {M.~D.}\
  \bibnamefont {Lukin}},\ }\bibfield  {title} {\bibinfo {title} {Parallel
  implementation of high-fidelity multiqubit gates with neutral atoms},\ }\href
  {https://doi.org/10.1103/PhysRevLett.123.170503} {\bibfield  {journal}
  {\bibinfo  {journal} {Phys. Rev. Lett.}\ }\textbf {\bibinfo {volume} {123}},\
  \bibinfo {pages} {170503} (\bibinfo {year} {2019})}\BibitemShut {NoStop}%
\bibitem [{\citenamefont {Graham}\ \emph {et~al.}(2019)\citenamefont {Graham},
  \citenamefont {Kwon}, \citenamefont {Grinkemeyer}, \citenamefont {Marra},
  \citenamefont {Jiang}, \citenamefont {Lichtman}, \citenamefont {Sun},
  \citenamefont {Ebert},\ and\ \citenamefont {Saffman}}]{Graham2019}%
  \BibitemOpen
  \bibfield  {author} {\bibinfo {author} {\bibfnamefont {T.~M.}\ \bibnamefont
  {Graham}}, \bibinfo {author} {\bibfnamefont {M.}~\bibnamefont {Kwon}},
  \bibinfo {author} {\bibfnamefont {B.}~\bibnamefont {Grinkemeyer}}, \bibinfo
  {author} {\bibfnamefont {Z.}~\bibnamefont {Marra}}, \bibinfo {author}
  {\bibfnamefont {X.}~\bibnamefont {Jiang}}, \bibinfo {author} {\bibfnamefont
  {M.~T.}\ \bibnamefont {Lichtman}}, \bibinfo {author} {\bibfnamefont
  {Y.}~\bibnamefont {Sun}}, \bibinfo {author} {\bibfnamefont {M.}~\bibnamefont
  {Ebert}},\ and\ \bibinfo {author} {\bibfnamefont {M.}~\bibnamefont
  {Saffman}},\ }\bibfield  {title} {\bibinfo {title} {Rydberg-{{Mediated
  Entanglement}} in a {{Two-Dimensional Neutral Atom Qubit Array}}},\ }\href
  {https://doi.org/10.1103/PhysRevLett.123.230501} {\bibfield  {journal}
  {\bibinfo  {journal} {Phys. Rev. Lett.}\ }\textbf {\bibinfo {volume} {123}},\
  \bibinfo {pages} {230501} (\bibinfo {year} {2019})}\BibitemShut {NoStop}%
\bibitem [{\citenamefont {Madjarov}\ \emph {et~al.}(2020)\citenamefont
  {Madjarov}, \citenamefont {Covey}, \citenamefont {Shaw}, \citenamefont
  {Choi}, \citenamefont {Kale}, \citenamefont {Cooper}, \citenamefont
  {Pichler}, \citenamefont {Schkolnik}, \citenamefont {Williams},\ and\
  \citenamefont {Endres}}]{Madjarov2020}%
  \BibitemOpen
  \bibfield  {author} {\bibinfo {author} {\bibfnamefont {I.~S.}\ \bibnamefont
  {Madjarov}}, \bibinfo {author} {\bibfnamefont {J.~P.}\ \bibnamefont {Covey}},
  \bibinfo {author} {\bibfnamefont {A.~L.}\ \bibnamefont {Shaw}}, \bibinfo
  {author} {\bibfnamefont {J.}~\bibnamefont {Choi}}, \bibinfo {author}
  {\bibfnamefont {A.}~\bibnamefont {Kale}}, \bibinfo {author} {\bibfnamefont
  {A.}~\bibnamefont {Cooper}}, \bibinfo {author} {\bibfnamefont
  {H.}~\bibnamefont {Pichler}}, \bibinfo {author} {\bibfnamefont
  {V.}~\bibnamefont {Schkolnik}}, \bibinfo {author} {\bibfnamefont {J.~R.}\
  \bibnamefont {Williams}},\ and\ \bibinfo {author} {\bibfnamefont
  {M.}~\bibnamefont {Endres}},\ }\bibfield  {title} {\bibinfo {title}
  {High-fidelity entanglement and detection of alkaline-earth {{Rydberg}}
  atoms},\ }\href {https://doi.org/10.1038/s41567-020-0903-z} {\bibfield
  {journal} {\bibinfo  {journal} {Nat. Phys.}\ }\textbf {\bibinfo {volume}
  {16}},\ \bibinfo {pages} {857} (\bibinfo {year} {2020})}\BibitemShut
  {NoStop}%
\bibitem [{\citenamefont {Schine}\ \emph {et~al.}(2022)\citenamefont {Schine},
  \citenamefont {Young}, \citenamefont {Eckner}, \citenamefont {Martin},\ and\
  \citenamefont {Kaufman}}]{Schine2022}%
  \BibitemOpen
  \bibfield  {author} {\bibinfo {author} {\bibfnamefont {N.}~\bibnamefont
  {Schine}}, \bibinfo {author} {\bibfnamefont {A.~W.}\ \bibnamefont {Young}},
  \bibinfo {author} {\bibfnamefont {W.~J.}\ \bibnamefont {Eckner}}, \bibinfo
  {author} {\bibfnamefont {M.~J.}\ \bibnamefont {Martin}},\ and\ \bibinfo
  {author} {\bibfnamefont {A.~M.}\ \bibnamefont {Kaufman}},\ }\bibfield
  {title} {\bibinfo {title} {Long-lived {{Bell}} states in an array of optical
  clock qubits},\ }\href {https://doi.org/10.1038/s41567-022-01678-w}
  {\bibfield  {journal} {\bibinfo  {journal} {Nat. Phys.}\ }\textbf {\bibinfo
  {volume} {18}},\ \bibinfo {pages} {1067} (\bibinfo {year}
  {2022})}\BibitemShut {NoStop}%
\bibitem [{\citenamefont {Ma}\ \emph {et~al.}(2023)\citenamefont {Ma},
  \citenamefont {Liu}, \citenamefont {Peng}, \citenamefont {Zhang},
  \citenamefont {Jandura}, \citenamefont {Claes}, \citenamefont {Burgers},
  \citenamefont {Pupillo}, \citenamefont {Puri},\ and\ \citenamefont
  {Thompson}}]{Ma2023}%
  \BibitemOpen
  \bibfield  {author} {\bibinfo {author} {\bibfnamefont {S.}~\bibnamefont
  {Ma}}, \bibinfo {author} {\bibfnamefont {G.}~\bibnamefont {Liu}}, \bibinfo
  {author} {\bibfnamefont {P.}~\bibnamefont {Peng}}, \bibinfo {author}
  {\bibfnamefont {B.}~\bibnamefont {Zhang}}, \bibinfo {author} {\bibfnamefont
  {S.}~\bibnamefont {Jandura}}, \bibinfo {author} {\bibfnamefont
  {J.}~\bibnamefont {Claes}}, \bibinfo {author} {\bibfnamefont {A.~P.}\
  \bibnamefont {Burgers}}, \bibinfo {author} {\bibfnamefont {G.}~\bibnamefont
  {Pupillo}}, \bibinfo {author} {\bibfnamefont {S.}~\bibnamefont {Puri}},\ and\
  \bibinfo {author} {\bibfnamefont {J.~D.}\ \bibnamefont {Thompson}},\
  }\bibfield  {title} {\bibinfo {title} {High-fidelity gates and mid-circuit
  erasure conversion in an atomic qubit},\ }\href
  {https://doi.org/10.1038/s41586-023-06438-1} {\bibfield  {journal} {\bibinfo
  {journal} {Nature}\ }\textbf {\bibinfo {volume} {622}},\ \bibinfo {pages}
  {279} (\bibinfo {year} {2023})}\BibitemShut {NoStop}%
\bibitem [{\citenamefont {Bluvstein}\ \emph {et~al.}(2023)\citenamefont
  {Bluvstein}, \citenamefont {Evered}, \citenamefont {Geim}, \citenamefont
  {Li}, \citenamefont {Zhou}, \citenamefont {Manovitz}, \citenamefont {Ebadi},
  \citenamefont {Cain}, \citenamefont {Kalinowski}, \citenamefont {Hangleiter},
  \citenamefont {Ataides}, \citenamefont {Maskara}, \citenamefont {Cong},
  \citenamefont {Gao}, \citenamefont {Rodriguez}, \citenamefont {Karolyshyn},
  \citenamefont {Semeghini}, \citenamefont {Gullans}, \citenamefont {Greiner},
  \citenamefont {Vuletic},\ and\ \citenamefont {Lukin}}]{Bluvstein2023}%
  \BibitemOpen
  \bibfield  {author} {\bibinfo {author} {\bibfnamefont {D.}~\bibnamefont
  {Bluvstein}}, \bibinfo {author} {\bibfnamefont {S.~J.}\ \bibnamefont
  {Evered}}, \bibinfo {author} {\bibfnamefont {A.~A.}\ \bibnamefont {Geim}},
  \bibinfo {author} {\bibfnamefont {S.~H.}\ \bibnamefont {Li}}, \bibinfo
  {author} {\bibfnamefont {H.}~\bibnamefont {Zhou}}, \bibinfo {author}
  {\bibfnamefont {T.}~\bibnamefont {Manovitz}}, \bibinfo {author}
  {\bibfnamefont {S.}~\bibnamefont {Ebadi}}, \bibinfo {author} {\bibfnamefont
  {M.}~\bibnamefont {Cain}}, \bibinfo {author} {\bibfnamefont {M.}~\bibnamefont
  {Kalinowski}}, \bibinfo {author} {\bibfnamefont {D.}~\bibnamefont
  {Hangleiter}}, \bibinfo {author} {\bibfnamefont {J.~P.~B.}\ \bibnamefont
  {Ataides}}, \bibinfo {author} {\bibfnamefont {N.}~\bibnamefont {Maskara}},
  \bibinfo {author} {\bibfnamefont {I.}~\bibnamefont {Cong}}, \bibinfo {author}
  {\bibfnamefont {X.}~\bibnamefont {Gao}}, \bibinfo {author} {\bibfnamefont
  {P.~S.}\ \bibnamefont {Rodriguez}}, \bibinfo {author} {\bibfnamefont
  {T.}~\bibnamefont {Karolyshyn}}, \bibinfo {author} {\bibfnamefont
  {G.}~\bibnamefont {Semeghini}}, \bibinfo {author} {\bibfnamefont {M.~J.}\
  \bibnamefont {Gullans}}, \bibinfo {author} {\bibfnamefont {M.}~\bibnamefont
  {Greiner}}, \bibinfo {author} {\bibfnamefont {V.}~\bibnamefont {Vuletic}},\
  and\ \bibinfo {author} {\bibfnamefont {M.~D.}\ \bibnamefont {Lukin}},\
  }\href@noop {} {\bibinfo {title} {Logical quantum processor based on
  reconfigurable atom arrays}} (\bibinfo {year} {2023}),\ \Eprint
  {https://arxiv.org/abs/2312.03982} {arxiv:2312.03982} \BibitemShut {NoStop}%
\bibitem [{\citenamefont {Holland}\ \emph {et~al.}(2023)\citenamefont
  {Holland}, \citenamefont {Lu},\ and\ \citenamefont {Cheuk}}]{Holland2023}%
  \BibitemOpen
  \bibfield  {author} {\bibinfo {author} {\bibfnamefont {C.~M.}\ \bibnamefont
  {Holland}}, \bibinfo {author} {\bibfnamefont {Y.}~\bibnamefont {Lu}},\ and\
  \bibinfo {author} {\bibfnamefont {L.~W.}\ \bibnamefont {Cheuk}},\ }\bibfield
  {title} {\bibinfo {title} {On-demand entanglement of molecules in a
  reconfigurable optical tweezer array},\ }\href
  {https://doi.org/10.1126/science.adf4272} {\bibfield  {journal} {\bibinfo
  {journal} {Science}\ }\textbf {\bibinfo {volume} {382}},\ \bibinfo {pages}
  {1143} (\bibinfo {year} {2023})}\BibitemShut {NoStop}%
\bibitem [{\citenamefont {Bao}\ \emph {et~al.}(2023)\citenamefont {Bao},
  \citenamefont {Yu}, \citenamefont {Anderegg}, \citenamefont {Chae},
  \citenamefont {Ketterle}, \citenamefont {Ni},\ and\ \citenamefont
  {Doyle}}]{Bao2023}%
  \BibitemOpen
  \bibfield  {author} {\bibinfo {author} {\bibfnamefont {Y.}~\bibnamefont
  {Bao}}, \bibinfo {author} {\bibfnamefont {S.~S.}\ \bibnamefont {Yu}},
  \bibinfo {author} {\bibfnamefont {L.}~\bibnamefont {Anderegg}}, \bibinfo
  {author} {\bibfnamefont {E.}~\bibnamefont {Chae}}, \bibinfo {author}
  {\bibfnamefont {W.}~\bibnamefont {Ketterle}}, \bibinfo {author}
  {\bibfnamefont {K.-K.}\ \bibnamefont {Ni}},\ and\ \bibinfo {author}
  {\bibfnamefont {J.~M.}\ \bibnamefont {Doyle}},\ }\bibfield  {title} {\bibinfo
  {title} {Dipolar spin-exchange and entanglement between molecules in an
  optical tweezer array},\ }\href {https://doi.org/10.1126/science.adf8999}
  {\bibfield  {journal} {\bibinfo  {journal} {Science}\ }\textbf {\bibinfo
  {volume} {382}},\ \bibinfo {pages} {1138} (\bibinfo {year}
  {2023})}\BibitemShut {NoStop}%
\bibitem [{\citenamefont {Chen}\ \emph {et~al.}(2023)\citenamefont {Chen},
  \citenamefont {Lucas},\ and\ \citenamefont {Yin}}]{Chen2023}%
  \BibitemOpen
  \bibfield  {author} {\bibinfo {author} {\bibfnamefont {C.-F.~A.}\
  \bibnamefont {Chen}}, \bibinfo {author} {\bibfnamefont {A.}~\bibnamefont
  {Lucas}},\ and\ \bibinfo {author} {\bibfnamefont {C.}~\bibnamefont {Yin}},\
  }\bibfield  {title} {\bibinfo {title} {Speed limits and locality in many-body
  quantum dynamics},\ }\href {https://doi.org/10.1088/1361-6633/acfaae}
  {\bibfield  {journal} {\bibinfo  {journal} {Rep. Prog. Phys.}\ }\textbf
  {\bibinfo {volume} {86}},\ \bibinfo {pages} {116001} (\bibinfo {year}
  {2023})}\BibitemShut {NoStop}%
\bibitem [{\citenamefont {Eldredge}\ \emph {et~al.}(2017)\citenamefont
  {Eldredge}, \citenamefont {Gong}, \citenamefont {Young}, \citenamefont
  {Moosavian}, \citenamefont {{Foss-Feig}},\ and\ \citenamefont
  {Gorshkov}}]{Eldredge2017}%
  \BibitemOpen
  \bibfield  {author} {\bibinfo {author} {\bibfnamefont {Z.}~\bibnamefont
  {Eldredge}}, \bibinfo {author} {\bibfnamefont {Z.~X.}\ \bibnamefont {Gong}},
  \bibinfo {author} {\bibfnamefont {J.~T.}\ \bibnamefont {Young}}, \bibinfo
  {author} {\bibfnamefont {A.~H.}\ \bibnamefont {Moosavian}}, \bibinfo {author}
  {\bibfnamefont {M.}~\bibnamefont {{Foss-Feig}}},\ and\ \bibinfo {author}
  {\bibfnamefont {A.~V.}\ \bibnamefont {Gorshkov}},\ }\bibfield  {title}
  {\bibinfo {title} {Fast quantum state transfer and entanglement
  renormalization using long-range interactions},\ }\href
  {https://doi.org/10.1103/PhysRevLett.119.170503} {\bibfield  {journal}
  {\bibinfo  {journal} {Phys. Rev. Lett.}\ }\textbf {\bibinfo {volume} {119}},\
  \bibinfo {pages} {170503} (\bibinfo {year} {2017})}\BibitemShut {NoStop}%
\bibitem [{\citenamefont {Tran}\ \emph
  {et~al.}(2021{\natexlab{a}})\citenamefont {Tran}, \citenamefont {Guo},
  \citenamefont {Deshpande}, \citenamefont {Lucas},\ and\ \citenamefont
  {Gorshkov}}]{Tran2021a}%
  \BibitemOpen
  \bibfield  {author} {\bibinfo {author} {\bibfnamefont {M.~C.}\ \bibnamefont
  {Tran}}, \bibinfo {author} {\bibfnamefont {A.~Y.}\ \bibnamefont {Guo}},
  \bibinfo {author} {\bibfnamefont {A.}~\bibnamefont {Deshpande}}, \bibinfo
  {author} {\bibfnamefont {A.}~\bibnamefont {Lucas}},\ and\ \bibinfo {author}
  {\bibfnamefont {A.~V.}\ \bibnamefont {Gorshkov}},\ }\bibfield  {title}
  {\bibinfo {title} {Optimal state transfer and entanglement generation in
  power-law interacting systems},\ }\href
  {https://doi.org/10.1103/PhysRevX.11.031016} {\bibfield  {journal} {\bibinfo
  {journal} {Phys. Rev. X}\ }\textbf {\bibinfo {volume} {11}},\ \bibinfo
  {pages} {031016} (\bibinfo {year} {2021}{\natexlab{a}})}\BibitemShut
  {NoStop}%
\bibitem [{\citenamefont {Hong}\ and\ \citenamefont {Lucas}(2021)}]{Hong2021}%
  \BibitemOpen
  \bibfield  {author} {\bibinfo {author} {\bibfnamefont {Y.}~\bibnamefont
  {Hong}}\ and\ \bibinfo {author} {\bibfnamefont {A.}~\bibnamefont {Lucas}},\
  }\bibfield  {title} {\bibinfo {title} {Fast high-fidelity multiqubit state
  transfer with long-range interactions},\ }\href
  {https://doi.org/10.1103/PhysRevA.103.042425} {\bibfield  {journal} {\bibinfo
   {journal} {Phys. Rev. A}\ }\textbf {\bibinfo {volume} {103}},\ \bibinfo
  {pages} {042425} (\bibinfo {year} {2021})}\BibitemShut {NoStop}%
\bibitem [{\citenamefont {Bapat}\ \emph {et~al.}(2023)\citenamefont {Bapat},
  \citenamefont {Childs}, \citenamefont {Gorshkov},\ and\ \citenamefont
  {Schoute}}]{Bapat2023}%
  \BibitemOpen
  \bibfield  {author} {\bibinfo {author} {\bibfnamefont {A.}~\bibnamefont
  {Bapat}}, \bibinfo {author} {\bibfnamefont {A.~M.}\ \bibnamefont {Childs}},
  \bibinfo {author} {\bibfnamefont {A.~V.}\ \bibnamefont {Gorshkov}},\ and\
  \bibinfo {author} {\bibfnamefont {E.}~\bibnamefont {Schoute}},\ }\bibfield
  {title} {\bibinfo {title} {Advantages and {{Limitations}} of {{Quantum
  Routing}}},\ }\href {https://doi.org/10.1103/PRXQuantum.4.010313} {\bibfield
  {journal} {\bibinfo  {journal} {PRX Quantum}\ }\textbf {\bibinfo {volume}
  {4}},\ \bibinfo {pages} {010313} (\bibinfo {year} {2023})}\BibitemShut
  {NoStop}%
\bibitem [{\citenamefont {Bollinger}\ \emph {et~al.}(1996)\citenamefont
  {Bollinger}, \citenamefont {Itano}, \citenamefont {Wineland},\ and\
  \citenamefont {Heinzen}}]{Bollinger1996}%
  \BibitemOpen
  \bibfield  {author} {\bibinfo {author} {\bibfnamefont {J.~J.}\ \bibnamefont
  {Bollinger}}, \bibinfo {author} {\bibfnamefont {W.~M.}\ \bibnamefont
  {Itano}}, \bibinfo {author} {\bibfnamefont {D.~J.}\ \bibnamefont
  {Wineland}},\ and\ \bibinfo {author} {\bibfnamefont {D.~J.}\ \bibnamefont
  {Heinzen}},\ }\bibfield  {title} {\bibinfo {title} {Optimal frequency
  measurements with maximally correlated states},\ }\href
  {https://doi.org/10.1103/PhysRevA.54.R4649} {\bibfield  {journal} {\bibinfo
  {journal} {Phys. Rev. A}\ }\textbf {\bibinfo {volume} {54}},\ \bibinfo
  {pages} {R4649} (\bibinfo {year} {1996})}\BibitemShut {NoStop}%
\bibitem [{\citenamefont {Eldredge}\ \emph {et~al.}(2018)\citenamefont
  {Eldredge}, \citenamefont {{Foss-Feig}}, \citenamefont {Gross}, \citenamefont
  {Rolston},\ and\ \citenamefont {Gorshkov}}]{Eldredge2018}%
  \BibitemOpen
  \bibfield  {author} {\bibinfo {author} {\bibfnamefont {Z.}~\bibnamefont
  {Eldredge}}, \bibinfo {author} {\bibfnamefont {M.}~\bibnamefont
  {{Foss-Feig}}}, \bibinfo {author} {\bibfnamefont {J.~A.}\ \bibnamefont
  {Gross}}, \bibinfo {author} {\bibfnamefont {S.~L.}\ \bibnamefont {Rolston}},\
  and\ \bibinfo {author} {\bibfnamefont {A.~V.}\ \bibnamefont {Gorshkov}},\
  }\bibfield  {title} {\bibinfo {title} {Optimal and secure measurement
  protocols for quantum sensor networks},\ }\href
  {https://doi.org/10.1103/PhysRevA.97.042337} {\bibfield  {journal} {\bibinfo
  {journal} {Phys. Rev. A}\ }\textbf {\bibinfo {volume} {97}},\ \bibinfo
  {pages} {042337} (\bibinfo {year} {2018})}\BibitemShut {NoStop}%
\bibitem [{\citenamefont {Guo}\ \emph {et~al.}(2022)\citenamefont {Guo},
  \citenamefont {Deshpande}, \citenamefont {Chu}, \citenamefont {Eldredge},
  \citenamefont {Bienias}, \citenamefont {Devulapalli}, \citenamefont {Su},
  \citenamefont {Childs},\ and\ \citenamefont {Gorshkov}}]{Guo2022}%
  \BibitemOpen
  \bibfield  {author} {\bibinfo {author} {\bibfnamefont {A.~Y.}\ \bibnamefont
  {Guo}}, \bibinfo {author} {\bibfnamefont {A.}~\bibnamefont {Deshpande}},
  \bibinfo {author} {\bibfnamefont {S.-K.}\ \bibnamefont {Chu}}, \bibinfo
  {author} {\bibfnamefont {Z.}~\bibnamefont {Eldredge}}, \bibinfo {author}
  {\bibfnamefont {P.}~\bibnamefont {Bienias}}, \bibinfo {author} {\bibfnamefont
  {D.}~\bibnamefont {Devulapalli}}, \bibinfo {author} {\bibfnamefont
  {Y.}~\bibnamefont {Su}}, \bibinfo {author} {\bibfnamefont {A.~M.}\
  \bibnamefont {Childs}},\ and\ \bibinfo {author} {\bibfnamefont {A.~V.}\
  \bibnamefont {Gorshkov}},\ }\bibfield  {title} {\bibinfo {title}
  {Implementing a fast unbounded quantum fanout gate using power-law
  interactions},\ }\href {https://doi.org/10.1103/PhysRevResearch.4.L042016}
  {\bibfield  {journal} {\bibinfo  {journal} {Phys. Rev. Research}\ }\textbf
  {\bibinfo {volume} {4}},\ \bibinfo {pages} {L042016} (\bibinfo {year}
  {2022})}\BibitemShut {NoStop}%
\bibitem [{\citenamefont {Hastings}\ and\ \citenamefont
  {Koma}(2006)}]{Hastings2006}%
  \BibitemOpen
  \bibfield  {author} {\bibinfo {author} {\bibfnamefont {M.~B.}\ \bibnamefont
  {Hastings}}\ and\ \bibinfo {author} {\bibfnamefont {T.}~\bibnamefont
  {Koma}},\ }\bibfield  {title} {\bibinfo {title} {Spectral gap and exponential
  decay of correlations},\ }\href {https://doi.org/10.1007/s00220-006-0030-4}
  {\bibfield  {journal} {\bibinfo  {journal} {Comm. Math. Phys.}\ }\textbf
  {\bibinfo {volume} {265}},\ \bibinfo {pages} {781} (\bibinfo {year}
  {2006})}\BibitemShut {NoStop}%
\bibitem [{\citenamefont {Guo}\ \emph {et~al.}(2020)\citenamefont {Guo},
  \citenamefont {Tran}, \citenamefont {Childs}, \citenamefont {Gorshkov},\ and\
  \citenamefont {Gong}}]{Guo2020}%
  \BibitemOpen
  \bibfield  {author} {\bibinfo {author} {\bibfnamefont {A.~Y.}\ \bibnamefont
  {Guo}}, \bibinfo {author} {\bibfnamefont {M.~C.}\ \bibnamefont {Tran}},
  \bibinfo {author} {\bibfnamefont {A.~M.}\ \bibnamefont {Childs}}, \bibinfo
  {author} {\bibfnamefont {A.~V.}\ \bibnamefont {Gorshkov}},\ and\ \bibinfo
  {author} {\bibfnamefont {Z.-X.}\ \bibnamefont {Gong}},\ }\bibfield  {title}
  {\bibinfo {title} {Signaling and {{Scrambling}} with {{Strongly Long-Range
  Interactions}}},\ }\href {https://doi.org/10.1103/PhysRevA.102.010401}
  {\bibfield  {journal} {\bibinfo  {journal} {Phys. Rev. A}\ }\textbf {\bibinfo
  {volume} {102}},\ \bibinfo {pages} {010401} (\bibinfo {year}
  {2020})}\BibitemShut {NoStop}%
\bibitem [{\citenamefont {Tran}\ \emph {et~al.}(2019)\citenamefont {Tran},
  \citenamefont {Guo}, \citenamefont {Su}, \citenamefont {Garrison},
  \citenamefont {Eldredge}, \citenamefont {{Foss-Feig}}, \citenamefont
  {Childs},\ and\ \citenamefont {Gorshkov}}]{Tran2019}%
  \BibitemOpen
  \bibfield  {author} {\bibinfo {author} {\bibfnamefont {M.~C.}\ \bibnamefont
  {Tran}}, \bibinfo {author} {\bibfnamefont {A.~Y.}\ \bibnamefont {Guo}},
  \bibinfo {author} {\bibfnamefont {Y.}~\bibnamefont {Su}}, \bibinfo {author}
  {\bibfnamefont {J.~R.}\ \bibnamefont {Garrison}}, \bibinfo {author}
  {\bibfnamefont {Z.}~\bibnamefont {Eldredge}}, \bibinfo {author}
  {\bibfnamefont {M.}~\bibnamefont {{Foss-Feig}}}, \bibinfo {author}
  {\bibfnamefont {A.~M.}\ \bibnamefont {Childs}},\ and\ \bibinfo {author}
  {\bibfnamefont {A.~V.}\ \bibnamefont {Gorshkov}},\ }\bibfield  {title}
  {\bibinfo {title} {Locality and {{Digital Quantum Simulation}} of {{Power-Law
  Interactions}}},\ }\href {https://doi.org/10.1103/PhysRevX.9.031006}
  {\bibfield  {journal} {\bibinfo  {journal} {Phys. Rev. X}\ }\textbf {\bibinfo
  {volume} {9}},\ \bibinfo {pages} {031006} (\bibinfo {year}
  {2019})}\BibitemShut {NoStop}%
\bibitem [{\citenamefont {Kuwahara}\ and\ \citenamefont
  {Saito}(2020)}]{kuwaharaStrictlyLinearLight2020}%
  \BibitemOpen
  \bibfield  {author} {\bibinfo {author} {\bibfnamefont {T.}~\bibnamefont
  {Kuwahara}}\ and\ \bibinfo {author} {\bibfnamefont {K.}~\bibnamefont
  {Saito}},\ }\bibfield  {title} {\bibinfo {title} {Strictly linear light cones
  in long-range interacting systems of arbitrary dimensions},\ }\href
  {https://doi.org/10.1103/PhysRevX.10.031010} {\bibfield  {journal} {\bibinfo
  {journal} {Phys. Rev. X}\ }\textbf {\bibinfo {volume} {10}},\ \bibinfo
  {pages} {031010} (\bibinfo {year} {2020})}\BibitemShut {NoStop}%
\bibitem [{\citenamefont {Tran}\ \emph
  {et~al.}(2021{\natexlab{b}})\citenamefont {Tran}, \citenamefont {Guo},
  \citenamefont {Baldwin}, \citenamefont {Ehrenberg}, \citenamefont
  {Gorshkov},\ and\ \citenamefont {Lucas}}]{Tran2021b}%
  \BibitemOpen
  \bibfield  {author} {\bibinfo {author} {\bibfnamefont {M.~C.}\ \bibnamefont
  {Tran}}, \bibinfo {author} {\bibfnamefont {A.~Y.}\ \bibnamefont {Guo}},
  \bibinfo {author} {\bibfnamefont {C.~L.}\ \bibnamefont {Baldwin}}, \bibinfo
  {author} {\bibfnamefont {A.}~\bibnamefont {Ehrenberg}}, \bibinfo {author}
  {\bibfnamefont {A.~V.}\ \bibnamefont {Gorshkov}},\ and\ \bibinfo {author}
  {\bibfnamefont {A.}~\bibnamefont {Lucas}},\ }\bibfield  {title} {\bibinfo
  {title} {Lieb-{{Robinson}} light cone for power-law interactions},\ }\href
  {https://doi.org/10.1103/PhysRevLett.127.160401} {\bibfield  {journal}
  {\bibinfo  {journal} {Phys. Rev. Lett.}\ }\textbf {\bibinfo {volume} {127}},\
  \bibinfo {pages} {160401} (\bibinfo {year} {2021}{\natexlab{b}})}\BibitemShut
  {NoStop}%
\bibitem [{\citenamefont {Richerme}\ \emph {et~al.}(2014)\citenamefont
  {Richerme}, \citenamefont {Gong}, \citenamefont {Lee}, \citenamefont {Senko},
  \citenamefont {Smith}, \citenamefont {{Foss-Feig}}, \citenamefont
  {Michalakis}, \citenamefont {Gorshkov},\ and\ \citenamefont
  {Monroe}}]{Richerme2014}%
  \BibitemOpen
  \bibfield  {author} {\bibinfo {author} {\bibfnamefont {P.}~\bibnamefont
  {Richerme}}, \bibinfo {author} {\bibfnamefont {Z.-X.}\ \bibnamefont {Gong}},
  \bibinfo {author} {\bibfnamefont {A.}~\bibnamefont {Lee}}, \bibinfo {author}
  {\bibfnamefont {C.}~\bibnamefont {Senko}}, \bibinfo {author} {\bibfnamefont
  {J.}~\bibnamefont {Smith}}, \bibinfo {author} {\bibfnamefont
  {M.}~\bibnamefont {{Foss-Feig}}}, \bibinfo {author} {\bibfnamefont
  {S.}~\bibnamefont {Michalakis}}, \bibinfo {author} {\bibfnamefont {A.~V.}\
  \bibnamefont {Gorshkov}},\ and\ \bibinfo {author} {\bibfnamefont
  {C.}~\bibnamefont {Monroe}},\ }\bibfield  {title} {\bibinfo {title}
  {Non-local propagation of correlations in quantum systems with long-range
  interactions},\ }\href {https://doi.org/10.1038/nature13450} {\bibfield
  {journal} {\bibinfo  {journal} {Nature}\ }\textbf {\bibinfo {volume} {511}},\
  \bibinfo {pages} {198} (\bibinfo {year} {2014})}\BibitemShut {NoStop}%
\bibitem [{\citenamefont {Young}\ \emph {et~al.}(2021)\citenamefont {Young},
  \citenamefont {Bienias}, \citenamefont {Belyansky}, \citenamefont {Kaufman},\
  and\ \citenamefont {Gorshkov}}]{Young2021}%
  \BibitemOpen
  \bibfield  {author} {\bibinfo {author} {\bibfnamefont {J.~T.}\ \bibnamefont
  {Young}}, \bibinfo {author} {\bibfnamefont {P.}~\bibnamefont {Bienias}},
  \bibinfo {author} {\bibfnamefont {R.}~\bibnamefont {Belyansky}}, \bibinfo
  {author} {\bibfnamefont {A.~M.}\ \bibnamefont {Kaufman}},\ and\ \bibinfo
  {author} {\bibfnamefont {A.~V.}\ \bibnamefont {Gorshkov}},\ }\bibfield
  {title} {\bibinfo {title} {Asymmetric {{Blockade}} and {{Multiqubit Gates}}
  via {{Dipole-Dipole Interactions}}},\ }\href
  {https://doi.org/10.1103/PhysRevLett.127.120501} {\bibfield  {journal}
  {\bibinfo  {journal} {Phys. Rev. Lett.}\ }\textbf {\bibinfo {volume} {127}},\
  \bibinfo {pages} {120501} (\bibinfo {year} {2021})}\BibitemShut {NoStop}%
\bibitem [{\citenamefont {Saffman}\ \emph {et~al.}(2010)\citenamefont
  {Saffman}, \citenamefont {Walker},\ and\ \citenamefont
  {M{\o}lmer}}]{Saffman2010}%
  \BibitemOpen
  \bibfield  {author} {\bibinfo {author} {\bibfnamefont {M.}~\bibnamefont
  {Saffman}}, \bibinfo {author} {\bibfnamefont {T.~G.}\ \bibnamefont
  {Walker}},\ and\ \bibinfo {author} {\bibfnamefont {K.}~\bibnamefont
  {M{\o}lmer}},\ }\bibfield  {title} {\bibinfo {title} {Quantum information
  with {{Rydberg}} atoms},\ }\href {https://doi.org/10.1103/RevModPhys.82.2313}
  {\bibfield  {journal} {\bibinfo  {journal} {Rev. Mod. Phys.}\ }\textbf
  {\bibinfo {volume} {82}},\ \bibinfo {pages} {2313} (\bibinfo {year}
  {2010})}\BibitemShut {NoStop}%
\bibitem [{\citenamefont {Shi}\ and\ \citenamefont {Kennedy}(2017)}]{Shi2017}%
  \BibitemOpen
  \bibfield  {author} {\bibinfo {author} {\bibfnamefont {X.-F.}\ \bibnamefont
  {Shi}}\ and\ \bibinfo {author} {\bibfnamefont {T.~A.~B.}\ \bibnamefont
  {Kennedy}},\ }\bibfield  {title} {\bibinfo {title} {Annulled van der
  {{Waals}} interaction and fast {{Rydberg}} quantum gates},\ }\href
  {https://doi.org/10.1103/PhysRevA.95.043429} {\bibfield  {journal} {\bibinfo
  {journal} {Phys. Rev. A}\ }\textbf {\bibinfo {volume} {95}},\ \bibinfo
  {pages} {043429} (\bibinfo {year} {2017})}\BibitemShut {NoStop}%
\bibitem [{\citenamefont {Belyansky}\ \emph {et~al.}(2019)\citenamefont
  {Belyansky}, \citenamefont {Young}, \citenamefont {Bienias}, \citenamefont
  {Eldredge}, \citenamefont {Kaufman}, \citenamefont {Zoller},\ and\
  \citenamefont {Gorshkov}}]{Belyansky2019}%
  \BibitemOpen
  \bibfield  {author} {\bibinfo {author} {\bibfnamefont {R.}~\bibnamefont
  {Belyansky}}, \bibinfo {author} {\bibfnamefont {J.~T.}\ \bibnamefont
  {Young}}, \bibinfo {author} {\bibfnamefont {P.}~\bibnamefont {Bienias}},
  \bibinfo {author} {\bibfnamefont {Z.}~\bibnamefont {Eldredge}}, \bibinfo
  {author} {\bibfnamefont {A.~M.}\ \bibnamefont {Kaufman}}, \bibinfo {author}
  {\bibfnamefont {P.}~\bibnamefont {Zoller}},\ and\ \bibinfo {author}
  {\bibfnamefont {A.~V.}\ \bibnamefont {Gorshkov}},\ }\bibfield  {title}
  {\bibinfo {title} {Nondestructive {{Cooling}} of an {{Atomic Quantum
  Register}} via {{State-Insensitive Rydberg Interactions}}},\ }\href
  {https://doi.org/10.1103/PhysRevLett.123.213603} {\bibfield  {journal}
  {\bibinfo  {journal} {Phys. Rev. Lett.}\ }\textbf {\bibinfo {volume} {123}},\
  \bibinfo {pages} {213603} (\bibinfo {year} {2019})}\BibitemShut {NoStop}%
\bibitem [{\citenamefont {Yao}\ \emph {et~al.}(2015)\citenamefont {Yao},
  \citenamefont {Bennett}, \citenamefont {Laumann}, \citenamefont {Lev},\ and\
  \citenamefont {Gorshkov}}]{Yao2015}%
  \BibitemOpen
  \bibfield  {author} {\bibinfo {author} {\bibfnamefont {N.~Y.}\ \bibnamefont
  {Yao}}, \bibinfo {author} {\bibfnamefont {S.~D.}\ \bibnamefont {Bennett}},
  \bibinfo {author} {\bibfnamefont {C.~R.}\ \bibnamefont {Laumann}}, \bibinfo
  {author} {\bibfnamefont {B.~L.}\ \bibnamefont {Lev}},\ and\ \bibinfo {author}
  {\bibfnamefont {A.~V.}\ \bibnamefont {Gorshkov}},\ }\bibfield  {title}
  {\bibinfo {title} {Bilayer fractional quantum {{Hall}} states with dipoles},\
  }\href {https://doi.org/10.1103/PhysRevA.92.033609} {\bibfield  {journal}
  {\bibinfo  {journal} {Phys. Rev. A.}\ }\textbf {\bibinfo {volume} {92}},\
  \bibinfo {pages} {033609} (\bibinfo {year} {2015})}\BibitemShut {NoStop}%
\bibitem [{Note1()}]{Note1}%
  \BibitemOpen
  \bibinfo {note} {Because only the total accumulated phase matters in choosing
  the evolution time, we also expect the protocol to be robust against
  experimental errors such as uncertainties in the positions of individual
  particles: If the position of each particle is known up to a precision
  $\varepsilon \ll 1$, the total worst-case error in the accumulated phase
  scales as $t({r_1^{2d}}/{r^\alpha })\times (\varepsilon /r)$, with $r_1$
  being the length of each hypercubes and $r$ being the minimum distance
  between them. The result is a relative phase error proportional to
  $\varepsilon /r$, which becomes smaller and smaller as the distance between
  the hypercubes increases. Moreover, we expect the relative error to be even
  smaller in the commonly occurring situation when uncertainties in the
  positions are uncorrelated between different particles}\BibitemShut {NoStop}%
\end{thebibliography}%
\end{document}